

\input phyzzx
\tolerance=3000
\def\piedepagina{
 \footline={\ifnum\pageno>1 \hss\tenrm\folio\hss
            \else \hfil \fi}}
\mathsurround=2pt

\piedepagina

%
\def\F{{\cal F}}

\def\cG{\cal G}
\def\cA{{\cal A}}
\def\cB{{\cal B}}
%
%
\def\mn{{\mu\nu}}
\def\a{\alpha}
\def\b{\beta}
\def\ga{\gamma}
\def\m{\mu}
\def\n{\nu}
\def\r{\rho}
\def\s{\sigma}
\def\l{\lambda}
\def\d{\delta}
\def\dd{\Delta}
\def\t{\tau}
\def\ee{\epsilon}
\def\eps{\varepsilon}
\def\gm{\Gamma}
\def\gmb{\bar{\gm}}
\def\gmbn{\bar{\gm}^{(n)}}
\def\gmbz{\bar{\gm}^{(0)}}
\def\gmbo{\bar{\gm}^{(1)}}
\def\gmbol{\bar{\gm}^{(1)}_{\rm local}}
\def\gmbt{\bar{\gm}^{(2)}}
\def\gmal{\gm(\a)}
\def\gmbe{\gm(\b)}
\def\gmga{\gm(\ga)}
%
%
\def\tp{\tilde{p}}
\def\hp{\hat{p}}
\def\tk{\tilde{k}}
\def\hk{\hat{k}}

\def\ird{\underline{\omega}}
\def\uvd{\overline{\omega}}
\def\WW{\scriptscriptstyle R}
\def\VV{\scriptscriptstyle V}
\def\RR{{\rm I\!\!\, R}}
\def\pid{{(4\pi)}^D}
\def\pidhalf{(4\pi)^{D/2}}
\def\dhalf{{D\over 2}}
\def\gmdhalf{\gm\!\left( \dhalf \right)}
\def\meas{\int {d^D\! k\over (2\pi)^D} \, {d^D\! q \over (2\pi)^D}\,\, }
\def\meask{\int {d^D\! k\over (2\pi)^D}\,\,}
\def\measq{\int {d^D\! q\over (2\pi)^D}\,\,}
\def\RR{{\rm I\!\!\, R}}
\def\cv{c_{\VV}}
\def\xx{{g^2 \cv \over {4\pi}}}
\def\yy{{\cv\over k}}
\def\idx{\int d^3\!x}
\def\idp{\int {{d^3\!p}\over {(2\pi)^3}}\,{\rm e}^{-ip(x-y)}}
\def\idpp{\int {{d^3\!p_1}\over {(2\pi)^3}}\,
              {{d^3\!p_2}\over {(2\pi)^3}}\,
              {\rm e}^{-ip_1(x-z)}\,{\rm e}^{-ip_2(y-z)} }
%
%

%
%
\def\dj{{{\d}\over {\d J_{\m}^a}}}
\def\dgmbj{{{\d\gmb}\over {\d J_{\m}^a}}}

%
%
\def\dG{{{\d}\over {\d G_{\m}^a}}}
\def\dgmbG{{{\d\gmb}\over {\d G_{\m}^a}}}
\def\dgmbzG{{{\d\gmbz}\over {\d G_{\m}^a}}}
\def\dgmboG{{{\d\gmbo}\over {\d G_{\m}^a}}}
\def\dgmbolG{{{\d\gmbol}\over {\d G_{\m}^a}}}
%

%
\def\da{{{\d}\over {\d A^{\m a}}}}
\def\dgmba{{{\d\gmb}\over {\d A^{\m a}}}}
\def\dgmbza{{{\d\gmbz}\over {\d A^{\m a}}}}
\def\dgmboa{{{\d\gmbo}\over {\d A^{\m a}}}}
\def\dgmbola{{{\d\gmbol}\over {\d A^{\m a}}}}
%
%
\def\dcb{{{\d}\over{\d \bar{c}^a}}}
%
%
\def\dc{{{\d}\over{\d c^a}}}
\def\dgmbc{{{\d\gmb}\over{\d c^a}}}
\def\dgmbzc{{{\d\gmbz}\over{\d c^a}}}
\def\dgmboc{{{\d\gmbo}\over{\d c^a}}}
\def\dgmbolc{{{\d\gmbol}\over{\d c^a}}}
%
%
\def\dH{{{\d}\over{\d H^a}}}
\def\dgmbH{{{\d\gmb}\over{\d H^a}}}
\def\dgmbzH{{{\d\gmbz}\over{\d H^a}}}
\def\dgmboH{{{\d\gmbo}\over{\d H^a}}}
\def\dgmbolH{{{\d\gmbol}\over{\d H^a}}}


{\rightline{DAMTP 91-34}}
{\rightline{LPTHE 91-61}}
{\rightline{NBI-HE 91-55}}
{\rightline{UGMS 91-23}}

\vskip 1.8 true cm

{\centerline{\fourteenbf Chern-Simons Theory as the Large Mass Limit}}
{\centerline{\fourteenbf of Topologically Massive Yang-Mills Theory}}
\vskip 0.9 true cm
\centerline{G. Giavarini}
\vskip 0.2 true cm
\centerline{\it DAMTP, University of Cambridge}
\centerline{\it Silver Street, Cambridge CB3 9EW, UK}
\centerline{\it and}
\centerline{\it Laboratoire de Physique Th\'eorique et Hautes Energies,
Universit\'e Paris VI-VII $^*$ }
\centerline{\it Tour 14-24, $5^{eme}$ \'etage,
2 Place Jussieu, 75251 Paris Cedex 05, France}
\vskip 0.9 true cm
\centerline{C. P. Martin}
\vskip 0.2 true cm
\centerline{\it DAMTP, University of Cambridge}
\centerline{\it Silver Street, Cambridge CB3 9EW, UK}
\centerline{\it and}
\centerline{\it Department of Mathematics and Statistics,
         University of Guelph\footnote *{\rm Present address.}}
\centerline{\it Guelph, Ontario N1G 2W1, Canada}
\vskip 0.9 true cm
\centerline{F. Ruiz Ruiz}
\vskip 0.2 true cm
\centerline{\it The Niels Bohr Institute, University of Copenhagen}
\centerline{\it Blegdamsvej 17, DK-2100 Copenhagen \O, Denmark}

\vfill\eject

\vglue 4 true cm

{\leftskip=1 true cm \rightskip=1 true cm
\centerline{{\fourteenbf Abstract}}
\medskip
\noindent
We study quantum Chern-Simons theory as the large mass limit of the
limit $D\to 3$ of dimensionally regularized topologically massive
Yang-Mills  theory. This approach can also be interpreted as a
BRS-invariant hybrid
regularization of Chern-Simons theory, consisting of a higher-covariant
derivative Yang-Mills term plus dimensional regularization.
Working in the Landau gauge, we compute radiative corrections
up to second order in perturbation theory and show that there is
no two-loop correction to the one-loop
shift $k\rightarrow k+\cv,\,\,k$ being the bare Chern-Simons
parameter. In passing we also prove by explicit
computation that topologically
massive Yang-Mills theory is UV finite.
\par
}

\vfil\eject

\noindent{\bf\chapter{Introduction}}

Some ten years ago it came to light that massive non-abelian gauge
theories can be constructed in three dimensions without breaking local
gauge invariance nor introducing in the classical lagrangian fields other
than the gauge field\REF\Templeton {R. Jackiw and S. Templeton, Phys. Rev.
{\bf D23} (1981) 2291.}\REF\Schonfeld{J. Schonfeld Nucl. Phys. {\bf B185}
(1981) 157.}[\Templeton,\Schonfeld]. A 3-dimensional alternative to the
Higgs and Schwinger mechanisms was thus discovered. Massive gauge
excitations were obtained by adding to the standard Yang-Mills action the
intrinsically 3-dimensional Chern-Simons (CS) term, \ie\ the integral over
the spacetime manifold of the CS 3-form. The resulting
gauge-invariant theory was named topologically massive Yang-Mills
(TMYM) theory\REF\Deser{S. Deser, R. Jackiw and S. Templeton,
Ann. Phys. {\bf 140} (1982) 372.}[\Deser]. The parameter, say $m,$ in front
of the CS term gives the bare mass of the gauge field, its square
being the pole of the propagator in Minskowski spacetime, and is sometimes
called topological mass. The CS term has some interesting properties.
On the one hand, it is invariant under local gauge
transformations connected to the identity and is a topological invariant.
On the other, it is neither time-reversal invariant nor
parity-preserving, and changes under topologically non-trivial gauge
transformations. Invariance of the path integral under topologically
non-trivial gauge transformations demands quantizing the ratio of the
bare mass and the square of the coupling constant\REF\Desert{S. Deser,
R. Jackiw and S. Templeton, Phys. Rev. Lett. {\bf 48} (1982)
975.}[\Desert].

Some of the perturbative properties of TMYM theory have been studied
in refs.\REF\Pisarski{R. Pisarski and S. Rao, Phys. Rev. {\bf D32} (1985)
2081.}[\Deser,\Pisarski]. In particular, UV finiteness for the theory has
been shown at the one-loop level and anticipated to occur at two loops
[\Deser] if gauge invariance is respected by the UV divergent
contributions from every two-loop diagram. This is however a non-trivial
issue and its solution requires either ({\it i}) a
regularization-independent analysis, ({\it ii}) a gauge-invariant
regularization, or ({\it iii}) an explicit computation. As an intermediate
step towards the final goal of this paper, we will show by following the
last two approaches that indeed no UV divergences arise at two loops and
beyond. TMYM theory will thus turn out to be UV finite, although
it is only superrenormalizable by power counting. In [\Pisarski] the
zero-momentum limit of 1PI Green functions with less than four external legs
was proved to exist at every order in the number of loops. The topological
CS term therefore improves the perturbative bad zero-momentum behaviour
[\Templeton] of the corresponding massless 3-dimensional Yang-Mills theory.

Two years ago a theory whose classical action is only the CS term was
proposed\REF\Witten{E. Witten, Commun. Math. Phys. {\bf 121} (1989)
351.}[\Witten]. The theory is called Chern-Simons theory and, being
metric-independent, is a particular instance of the so-called topological
field theories\REF\Wittentop{E. Witten, Commun. Math. Phys.
{\bf 17} (1988) 353.}[\Wittentop] (see\REF\Blau{D. Birmingham, M. Blau,
M. Rakowski and G. Thompson, {\it Topological field theory}, Phys. Rep.,
to appear.}[\Blau] for a recent review). Topological field theories lack
physical local degrees of freedom but have nonetheless applications to both
Mathematics and Physics [\Blau ]. In particular, CS theory provides, on the
physical side, a 3-dimensional setting for the understanding of 2-dimensional
conformal field theories \REF\Moore{G. Moore and N. Seiberg, Phys. Lett.
{\bf B220} (1989) 422.}[\Witten,\Moore].
It is also used in condensed matter physics, \eg\
in the study of heteroconjectures or 3-dimensional gases of non-relativistic
electrons\REF\Studer{J. Fr\"olich and U. Studer, {\it
$U(1)\times SU(2)\!$-gauge invariance of non-relativistic quantum
mechanics and generalized Hall effects}, ETH-TH/91-13 preprint.}[\Studer].
On the mathematical side, CS theory with gauge group $SU(N)$ gives a
3-dimensional, quantum field theoretical definition of the
Jones polynomial\REF\Jones{V.F.R Jones, Ann. Math. {\bf 126} (1987)
335.}[\Jones] of knot theory and its generalization to the
HOMFLY polynomial\REF\HOMFLY{P. Freyd, D. Yetter, J. Hoste, W.B.R. Lickorish,
K. Millet and O. Ocneanu, Bull. Am. Math. Soc. {\bf 12} (1985)
239.}[\HOMFLY]. These knot and link invariants
can be obtained from the vacuum expectations values of Wilson loops [\Witten].
It is important to remind that Wilson loops are both gauge-invariant
and metric-independent so that they are observables of the theory.

CS theory can be quantized non-perturbatively within the canonical
formalism, where exact results can be obtained\REF\Canonical{S.
Elitzur, G. Moore, A. Schwimmer and N. Seiberg, Nucl. Phys. {\bf B326}
(1989) 108.\subpar
T. Killingback, Phys. Lett. {\bf B219} (1989) 448.\subpar
G.V. Dunne, R. Jackiw and C.A. Trugenberger, Ann. Phys. {\bf 194} (1989)
197.\subpar
M. Bos and V.P. Nair, Int. J. Mod. Phys. {\bf A5} (1990)959.\subpar
J.M.F. Labastida, P.M. Llatas and A.V. Ramallo, Nucl. Phys. {\bf B348}
(1991) 651.}[\Canonical] (see \REF\Froehlich{J. Fr\"ohlich and C. King,
Commun. Math. Phys. {\bf 126} (1989) 167.}[\Froehlich] for another
non-perturbative approach). This fact does not make the perturbative
analysis to be an academic exercise. On the contrary, since many results
in the canonical formulation rely on Feynman path integrals, which
are not completely well-defined mathematical objects, a check
using techniques from perturbative field theory is
in order. Of outstanding importance is to see whether it is possible
to recover in perturbation theory the find
[\Witten] that the exact result for the vacuum expectation value of a
Wilson loop is a function of $q={\rm exp}\left( {2\pi i\over k+N}\right),$
where the bare CS parameter $k\,(>\!0)$ has been shifted to $k+N.$
More precisely, one would like to see how it occurs in the perturbative
framework the fact\REF\AlvarezGtwo{L. Alvarez-Gaum\'e, J.M.F.
Labastida and A.V. Ramallo, Nucl. Phys. B,
Proc. Suppl. {\bf 18B} (1990) 1.}[\AlvarezGtwo] that the monodromy
parameter of the theory and the central extension
of the corresponding 2-dimensional current algebra are $k+N$ and $k$
respectively, $k$ being the bare CS parameter.
Furthermore, perturbative quantization of CS theory has led to
analytic expressions for the coefficients of the Jones polynomial
and its generalizations in terms of integrals involving the gauge
propagator\REF\Cotta{E. Guadagnini, M. Martellini and M. Mintchev,
Phys. Lett. {\bf 224} (1989) 489; Nucl. Phys. {\bf B330} (1990) 575.\subpar
P. Cotta-Ramusino, E. Guadagnini and M. Mintchev, Nucl. Phys. {\bf B330}
(1990) 577.}[\Cotta]. These coefficients are knot and link invariants
and their meaning and properties remain largely
unkown\REF\Kauffman{L. Kauffman, {\it On Knots}, Princeton University
Press (1987).}[\Kauffman]. Having explicit analytic expressions for
them may help to unravel their significance. This program has been started
in [\Cotta], where new relationships for the HOMFLY coefficients have been
obtained.

There are already a good many papers on perturbative CS
theory $\!\!\!\!\!\!\!\!\!\!\!\!\!\!\!\!\!\!\!\!
\!\!\!\!\!\!\!\!\!\!\!\!\!\!\!\!\!\!\!\!$
\REF\Guadagnini{E. Guadagnini, M. Martellini and M. Mintchev,
Phys. Lett. {\bf B227} (1989) 111.}\REF\AlvarezG{L. Alvarez-Gaum\'e,
J.M.F. Labastida and A.V. Ramallo, Nucl. Phys. {\bf B334} (1990) 103.}
\REF\Asorey{M. Asorey and F. Falceto, Phys. Lett. {\bf 241} (1990) 31.}
\REF\Carmelo{C.P. Martin, Phys. Lett. {\bf B241} (1990) 513.}
\REF\Semenoff{W. Chen, G.W. Semenoff and Yong-Shi Wu, Mod. Phys. Lett.
{\bf A5} (1990) 1833.}\REF\Birmingham{D. Birmingham, R. Kantwoski and
M. Rakowski, Phys. Lett. {\bf B251} (1990) 121.}\REF\CPM{C.P. Martin,
Phys. Lett. {\bf B246} (1991) 87.}
\REF\Ruso{G.P. Korchemsky, Mod. Phys. Lett. {\bf A6} (1990) 727.}
\REF\SUSY{D. Birmingham, M. Rakowski and G.
Thompson, Nucl. Phys. {\bf B234} (1990) 367.\subpar
F. Delduc, F. Gieres and S.P. Sorella, Phys. Lett. {\bf B225} (1989)
367.\subpar
P.H. Damgaard and V.O. Rivelles, Phys. Lett. {\bf B245} (1989) 48.\subpar
C. Lucchesi and O. Piguet, {\it Supersymmetry of the Chern-Simons
theory on a general three-dimensional manifold}, MPI-PAE/Pth 20/91 and
UGVA-DPT 1991/04-713 preprints.}
\REF\Blasi{ A. Blasi and R. Collina,
Nucl. Phys. {\bf B345} (1990) 472.}
\REF\Delduc{F. Delduc, C, Lucchesi, O. Piguet and S.P. Sorella,
Nucl. Phys. {\bf B346} (1990) 513.}
\REF\Dorey{D. Daniel and N. Dorey, Phys. Lett {\bf B246} (1990) 82.\subpar
N. Dorey, Phys. Lett. {\bf B246} (1990) 87.}\REF\Shifman{M.A. Shifman,
Nucl. Phys. {\bf B352} (1991) 87.}
\REF\Natan{D. Bar-Natan, {\it
Perturbative Chern-Simons theory}, Princeton University preprint 1990.}
\REF\Leibbrandt{G. Leibbrandt and C.P. Martin, {\it Perturbative Chern-Simons
theory in the light-cone gauge}, CERN-TH.6180/91 preprint.}
\REF\Axelrod{S. Axelrod and I.M. Singer,
{\it Chern-Simons perturbation theory},
MIT preprint 1991.}
\REF\Alvarez{M. Alvarez and J.M.F. Labastida, {\it
Analysis of observables in Chern-Simons perturbation theory}, US-FT-10/91
preprint.}
\REF\GMR{G. Giavarini, C.P. Martin and F. Ruiz Ruiz, {\it
Supersymmetry breaking in a BRS-invariant regularization of Chern-Simons
theory}, LPTHE 91/49, NBI-HE 91-48 and UGMS 91/24
preprints.}
[\Cotta,\Guadagnini -- \GMR]. Most of them deal with the study of
the Landau-gauge effective action and its properties.
Although perturbative CS theory
is not UV finite by power counting but merely renormalizable,
both the beta function [\Blasi,\Delduc] and the anomalous dimensions of the
elementary fields [\Delduc] have been proved to vanish at any order in
perturbation theory, in accordance with non-perturbative results (see
also [\CPM,\Axelrod]). Several groups have undertaken the computation
of the one-loop effective action by using a number of regularization methods
[\Guadagnini-\Birmingham]. The picture that has emerged from these
calculations is that every BRS-invariant regularization method
[\Witten,\AlvarezG-\Carmelo] gives for the gauge-invariant part of the
one-loop effective action the CS term with coefficient
$k+N,\,\,k\,(>\!0)$ being the bare or classical CS parameter.
Thus, the bare parameter $k$ is shifted to
$k+N$ for all BRS-invariant regulators used as yet. On the
other hand, this shift fails to occur for regularization methods
that are not gauge-invariant [\Guadagnini,\CPM]. One-loop results suggest
therefore that BRS-invariant regulators will naturally provide
a monodromy parameter $k+N\,\,(k\!>\!0),$ as non-perturbtive results demand.
(We refer the reader to the
conclusions for a further discussion of this idea).
Of course, for a complete agreement with the hamiltonian formalism,
there must be no corrections to $k+N$ beyond first order in perturbation
theory [\Alvarez].

The main purpose of this paper is to show the absence of two-loop
corrections to the one-loop shift $k\to k+N$ for the BRS-invariant
regularization first introduced in [\Carmelo]. This regulator is a
hybrid method obtained by adding to the CS action a Yang-Mills term
and then, by regularizing the UV divergences that are left with
a BRS-invariant dimensional regularization. In this way the regularized
theory involves two regulating
parameters, one is the dimensionality $D$ of spacetime (we will also
use $\eps = D-3),$ and the other, which we call $m,$ is a cut-off
introduced by the Yang-Mills term. We shall define CS theory as the limit
$D \to 3,\,\, m\to \infty$ of the regularized theory. After
appropriate rescalings of the fields and coupling constant, the
completely regularized CS theory will turn out to be a dimensionally
regularized TMYM theory whose bare topological mass is $m,$ the former
Yang-Mills regulating parameter. Thus CS theory can be viewed as the large
mass limit of TMYM theory, provided the limit $D\to 3$ of dimensionally
regularized TMYM theory exits. We shall show that the latter limit
does exist (see Sects. 2 and 7 for a more precise discussion).

The layout of this paper is as follows. In Sect. 2 we introduce our
explicitly BRS-invariant regularization method for CS theory
and show how it can be interpreted as a dimensional regularization
of TMYM theory. Sect. 3 is devoted to some theorems that simplify
enormously the computation of the large mass limit of the integrals
arising from the Feynman diagrams we will have to consider. The
calculation of the 1PI Green functions needed to construct
the local part of the effective action up to two loops is presented in
Sects. 4, 5 and 6. For the sake of the reader, we collect in Sect. 7 the
main steps leading to our proving the UV finiteness for TMYM theory.
In Sect. 8 the complete local part of the bare effective action is obtained
up to second order in perturbation theory by using BRS techniques and the
values of the 1PI functions obtained in previous sections. The absence of
two-loop corrections to the one-loop shift $k \to k +N$ is discussed in
Sect. 9. Finally, Sect. 10 contains the conclusions we have drawn from
the analyis reported here. We also include an appendix
with the relevant integrals we have had to evaluate in
our computations.

\noindent {\bf\chapter{The large $m$ limit, Feynman rules and regularization}}

Let us consider $\RR^3$ with the euclidean metric and denote by
$\cA_{\m}=\cA_{\m}^a T^a$ the $SU(N)$ gauge connection in the
fundamental representation, $T^a$ being the antihermitean generators
of the Lie algebra in this representation. We take the structure
constants $f^{abc}$ of $SU(N)$ to be completely antisymmetric in their
indices, with $[T^a,T^b]=f^{abc}T^c.$ For the generators $T^a$ we choose
the normalization $\Tr (T^a T^b)={1\over 2}\,\d^{ab}.$ The corresponding
BRS-invariant action in the Landau gauge for
TMYM theory with bare mass $m>0$ and bare coupling constant $g_t$
is then given by [\Schonfeld,\Deser,\Pisarski ]
$$
\eqalign{
S_m=& -\,\,i\,m \int d^3\!x ~ \ee^{\m\r\n} \left( {1\over 2}\,\,
                                     \cA^a_\m \partial_\r \cA^a_\n +
      {1 \over 3!}\,g_{t}\,f^{abc}\cA^a_\m \cA^b_\r \cA^c_\n  \right) \cr
    & + {1\over 4}\int d^3\!x ~\F^a_{\mn}\F^a_{\mn}
         - \int d^3\!x ~\left(\cB^a\partial
              \cA^a-\bar{c}^a \partial {\cal D}^{ab} c^b \right) \,\, , \cr}
\eqn\TMYM
$$
where $\F^a_{\m\n}=\partial_\m \cA^a_\n - \partial_\n \cA^a_\m
+ g_{t}f^{abc}\cA^b_\m \cA^c_\n$ is the field strength and
${\cal D}^{ab}_{\m}= \partial_\m \d^{ab}+g_{t}f^{acb}\cA_\m^c$
the covariant derivative. The mass dimensions of the fields
$\cA^a_{\m},~\cB^a,~\bar{c}^a,~c^a$ and of the coupling
constant $g_t$ are respectively $1/2,~3/2,~0,~1$ and $1/2.$
As for $m,$ it has dimensions of mass.

In turn, the BRS-invariant action in the Landau gauge for
CS theory is defined to be
$$
\eqalign{
S=& -\,\,i\,\int d^3\!x ~ \ee^{\m\r\n} \left( {1\over 2}\,\,
                                      A^a_\m \partial_\r A^a_\n +
     {1 \over 3!}\,g\,f^{abc}A^a_\m A^b_\r A^c_\n  \right) \cr
  & - \int d^3\!x ~\left(b^a\partial A^a-\bar{c}^a
                         \partial  D^{ab} c^b \right) \,\, , \cr}
\eqn\CSact
$$
with $A^a_{\m}$ is the gauge field on $\RR^3$ and
$D^{ab}_{\m}=\partial_{\m}\delta^{ab}+gf^{acb}A^c_{\m}.$
The fields $A^a_{\m},~b^a,~\bar{c}^a$ and $c^a$ now have
mass dimensions $1,~1,~0$ and $1,$ respectively; the coupling
constant $g$ is dimensionless.

It has been suggested\REF\Jackiw{R. Jackiw, {\it Topics in Planar
Physics}, in {\it Integrability and quantization}, M. Asorey, J. F.
Cari\~nena and L. A. Ibort eds., Nucl. Phys. B, Proc. Suppl. {\bf 18A}
(1990) 107.}[\Jackiw] that the limit $m\to\infty$ of TMYM theory with
classical action \TMYM\ should yield pure CS theory with classical action
\CSact. The purpose of this section is to propose a way to define this
limit so that it gives the desired result, at least in the highly
non-trivial instances we have checked.

We begin by considering tree-level Green functions
for the field $\cA^a_{\m}.$ By taking into acount the Feynman rules
in ref. [\Pisarski] one immediately realizes that their large $m$ limit
vanishes. This is due to the fast oscillations as $m\to\infty$ of the
Boltzman factor in the path integral, owing to the CS term in $S_m.$
Thus, to obtain pure CS theory as the large $m$ limit of TMYM theory,
the limit $m\to \infty$ cannot be directly taken over Green functions
in terms of $g_t$ but some other way must be contrived. Notice that
the dimensions of the fields and coupling constant in $S_m$ in eq.
\TMYM\ do not match those of the corresponding fields and coupling
constant in $S$ in eq. \CSact, being then little wonder that pure CS
theory is not obtained by taking the large $m$ limit as above.
To remedy this dimensional mismatch one introduces the following
scalings of the fields $\cA^a_{\m}$ and $\cB^a$ and the coupling
constant $g_t:$
$$
A^a_{\m}=m^{1/2}\cA^a_{\m}~~ , ~~~
b^a=m^{-1/2}\cB^a ~~ , ~~~ g=m^{-1/2} g_t~~.
\eqn\scalings
$$
We have used the same notation for the rescaled fields and coupling
constant as in eq. \CSact\ for reasons that will become clear in a
moment. Notice that $gA=g_t \cA$ and that the mass dimensions of
$A^a_{\m},$ $b^a$ and $g$ in eq. \scalings\ are the same
as those of the corresponding fields and coupling constant in the CS
action \CSact. Similar scalings have been introduced by other authors
in the context of  CS quantum mechanics\REF\Dunne{G. Dunne,
R. Jackiw and C.A. Trugenberger, Phys. Rev {\bf D 41} (1990)
661.}[\Dunne].

In terms of the new fields and coupling constant, the TMYM
action \TMYM\ can be recast in the form
$$
\eqalign{
S_m=& -\,i\idx ~ \ee^{\m\r\n} \left( {1\over 2}\,\, A^a_\m \partial_\r A^a_\n
      + {1 \over 3!}\,g\,f^{abc}A^a_\m A^b_\r A^c_\n  \right) \crr
& + {1\over 4m}\int d^3\!x ~F^a_{\mn}F^{a\,\mn}
      - \int d^3\!x ~s\left(\bar{c}^a\partial A^a\,+\,J^a A^a\,
                                        -\,H^a c^a \right) ~ , \cr}
\eqn\maction
$$
with $F^a_{\mn}= \partial_\m A^a_\n - \partial_\n
A^a_\m + gf^{abc}A^b_\m A^c_\n.$ With the purpose of computing the
effective action using BRS techniques we have introduced
external fields $J^a_{\m}$ and $H^a$ coupled to the
non-linear BRS transforms $sA^a_{\m}$ and $sc^a$ respectively.
The action on the fields of the BRS operator $s$ is:
$$
 sA_{\m}^a\,=\,\left( D_{\m}c \right) ^a ~, ~
    s\bar{c}^a\,=\,b^a ~, ~
      sc^a\,=-\,\,{1\over 2}\,g\,f^{abc}\,c^bc^c ~ , ~
        sb^a\,=\,0 ~, ~ sJ^a_{\m}\,=\,sH^a\,=\,0 ~,
\eqn\brstrans
$$
$D^{ab}_{\m}$ being the covariant derivative in terms of $A^a_\m$ and $g.$

The action $S_m$ is invariant under $s,$ since $s$ itself is nilpotent,
\ie\ $s^2=0,$ and the terms in $S_m$ that depend only on the gauge field
are invariant under infinitesimal gauge transformations for any
value of $g.$ This brings along a comment on the quantization of
$g$ as defined in eq. \scalings. It is well-known [\Deser]
that invariance of TMYM theory under gauge transformations with
non-zero winding number demands $g_t$ in \TMYM\ to be quantized
according to $4\pi m/ g_t^2=1,2,3,\ldots$ Consequently, $g$ in \scalings\
must satisfy
$$
{4\pi \over g^2}\equiv k=1,2,3,\ldots ~.
$$
These quantization conditions, however, require large values of the
fields to occur and therefore they do not appear in the perturbative
regime we shall consider in this paper.

The Feynman rules obtained from the action $S_m$ in eq. \maction\
are the following:

\item{a)} Propagators. The gauge field propagator is
$$
<A^a_\m (p)\, A^b_\n(-p)>\,=\,D^{ab}_{\mn}(p)=
 \,\,{\d^{ab}\, m \over p^2\,(p^2+m^2) }\,\,
   \left( m\, \ee_{\m\r\n}\,p^\r
          + p^2\, g_{\mn} - p_\m p_\n \right) ~ .
\eqn\gaugeprop
$$

\noindent For the ghost and auxiliary field we have:
$$
<c^a (p)\bar{c}^b(p)>\,\,=\,D^{ab}(p)=-\,{\d^{ab}\over p^2} ~ ,
\eqn\ghostprop
$$
$$
<b^a (p)A^b_\m(-p)>\,\,=\, D^{ab}_\m(p)=
             i \, \d^{ab} ~{p_{\m} \over p^2} ~ .
\eqn\auxprop
$$
\item{b)} Vertices. The three- and four-gauge vertices are
$$
V^{abc}_{\m\n\r}\, (p,q,r)=igf^{abc} \left\{  \ee_{\m\n\r} +
     {1\over m} \bigl[ (q-r)_\m \,g_{\n\r}+
         (r-p)_\n \,g_{\r\m} + (p-q)_\r \,g_{\mn}
              \bigr]\right\} ~ ,
$$
$$
\eqalign{
V^{abcd}_{\m\n\r\s} = -\, {g^2\over m}\, \Bigl[ \,& f^{abe}f^{cde}
    ( g_{\m\r}\,g_{\n\s} - g_{\n\r}\,g_{\m\s} ) +
          f^{cbe}f^{ade} ( g_{\m\r}\,g_{\n\s}
             - g_{\n\m}\,g_{\r\s} )  \cr
 & + f^{dbe}f^{cae} ( g_{\s\r}\,g_{\n\m}
         - g_{\n\r}\,g_{\m\s})\,\Bigr] ~ , \cr}
\eqn\gaugevert
$$

\noindent while the ghost vertex reads
$$
V^{abc}_\m(q)=-igf^{abc}q_\m ~ .
\eqn\ghostvert
$$
Finally, the gauge-ghost-external field vertex $J^a_{\m} A^b_{\n} c^c$ and
the ghost-ghost-external field vertex $H^a c^b c^c$ are
$$
-\,g\,f^{abc}\,g_{\m\n}\quad {\rm and} \quad g\,f^{abc}\,\, \eqn\extvert
$$
respectively.
The diagramatic representation of these rules is depicted
in Fig. 1.

Taking the large $m$ limit at fixed $g$ of \gaugeprop-\ghostvert\
one recovers the Feynman rules for pure CS theory with coupling
constant $g,$ as obtained directly from the action \CSact.
Hence, the definition of pure $SU(N)$ Chern-Simons theory
as the large $m$ limit of $SU(N)$ topologically massive Yang-Mills
theory arises naturally. In particular, we define the $E\!$-point
gauge field Green function
$G^{a_1\cdots a_E}_{\m_1\cdots \m_E}(p_1,\ldots ,p_E;g)$
for pure CS theory as the limit
$$
\eqalign{
G^{a_1 \cdots a_E}_{\m_1 \cdots \m_E}(p_1,\ldots ,p_E;g)&=
           \lim_{m\to\infty}
                 <A^a_{\m_1}(p_1)\cdots A^{a_E}_{\m_E}(p_E)>(g)  \cr
     &=\lim_{m\to\infty} m^{E/2}\,
       <\cA^a_{\m_1}(p_1)\cdots \cA^{a_E}_{\m_E}(p_E)>(g)~~ .\cr }
\eqn\mlimit
$$
Here we have explicitly written the dependence on $g$ to
remind that the Green functions are evaluated at fixed (but arbitrary)
$g,$ not to be confused with $g_t.$ In what follows we will omit this
fact in the notation. The external momenta
$p_1,\ldots , p_E$ are also arbitrary but lie in a bounded domain and
are non-exceptional. Similar definitions can be introduced for Green
functions involving auxiliary and/or ghost fields.
All these Green functions are defined as
formal series in powers of $g$ whose coefficients are obtained
using Feynman diagramatic techniques.

Let us emphasize that, as already shown, eq. \mlimit\ holds at the
tree-level and that our approach is to regard it as a definition
for CS Green functions to all orders in perturbation theory.
This paper is devoted to study the existence of the large $m$
limit up to two loops. In particular, following eq. \mlimit\ we will
explicitly construct the local two-loop effective action for pure
CS theory.

It should be noticed that the definition in eq. \mlimit\ is not yet
meaningful in perturbation theory since TMYM theory is not UV finite
by power counting. The superficial UV degree $\omega_m$ of a
1PI Feynman diagram with $E_g$ external gauge lines, $E_{gh}$ external
ghost lines, $V_3$ vertices with three gauge legs, $V_4$ vertices with
four gauge legs, $V_{gh}$ vertices of type $\bar{c}Ac,$ $V_J$
vertices of type $JAc$ and $V_H$ vertices of type $Hcc$ is
$$
\omega_{m} = 3-{1\over 2}(E_g + 2 E_{gh} + V_3 + V_{gh} + 2 V_4\,+\,
3\,V_J\,+\,2\,V_H )
\,\, . \eqn\uvdegree
$$
We thus see that only a finite number of 1PI diagrams have $\omega_m\geq 0$
so that TMYM theory is superrenormalizable. Let us spell out the 1PI
functions that are primitively UV divergent. We shall not consider
diagrams contributing to $<\!A^a_{\m}\!>,$ since as we will show in
Sect. 4, they vanish upon regularization owing to colour algebra.
Then, the only superficially UV divergent Green functions at one loop are
the gauge field two- and three-point functions and the ghost propagator,
for which  $\omega_m=1,~\omega_m=0$ and $\omega_m=0$ respectively.
At second order in perturbation theory, only 1PI diagrams contributing
to the gauge field propagator have non-negative $\omega_m,$
namely $\omega_m=0.$ Regarding higher loop 1PI diagrams, they are all
superficially UV convergent. To provide the superficially UV divergent
TMYM 1PI diagrams with a well-defined mathematical meaning
we will use dimensional regularization. Its implementation
in the case at hand is not straightforward due to the presence of the
Levi-Civita symbol and will be the subject of the next subsection.
In this paper we will show by explicit computation that, once the UV
divergences have been dimensionally regularized, the sum of all 1PI
Feynman diagrams at a given order in perturbation theory is finite
as $\eps\to 0,$ $\eps=D-3$ being the dimensional regulator.
This implies that there is no need of UV divergent counterterms to define
the effective action for TMYM theory and that TMYM Green functions
[entering in \eg\ the RHS of eq. \mlimit] can be understood as the
limit $D\to 3$ of the corresponding dimensionally regularized functions.

Let us point out that the action of eq. \maction\ can be thought of as
generating a higher-covariant derivative regularization
\REF\Faddeev{L.D. Faddeev and A.A. Slavnov, {\it Gauge fields:
introduction to quantum theory}. The Benjamin Cummings Publishing
Company (1980).}[\Faddeev ] of pure CS theory.  As usual, such
regularization is not enough as to regularize all 1PI diagrams and,
in addition, a second regularization must be implemented. For
computational purposes one may consider dimensional regularization.
Thus, our proposal to obtain pure CS theory as the large $m$ limit
of dimensioanlly regularized TMYM theory can be alternatively viewed
as a hybrid regularization method: higher-covariant derivatives plus
dimensional regularization. This method is characterized by two regulators:
$\eps$ and $m.$ The prescription, in accordance with eq. \mlimit ,
is to first send $\eps$ to zero and then
$m$ to infinity. Although there is no systematic study of
such hybrid regularization methods, it may be argued that this double
limit should yield a finite result, since the beta
function [\Blasi,\Delduc] and the anomalous
dimensions of the elementary fields [\Delduc] vanish for CS
theory in the Landau gauge. Our definition of the large $m$ limit would
then make sense at any order in perturbation theory.
We will see that our two-loop explicit computations are in agreement with
this general statement, so in some sense we will have checked
a particular case of hybrid regularization
in a non-trivial context. Notice that the vanishing of the anomalous
dimensions of the fields in both CS and TMYM theories implies that
that the na\"\i ve dimensional matchings in eq. \scalings\ are not
destroyed by radiative corrections, at least in the Landau gauge.

Since TMYM theory involves the
Levi-Civita tensor, the first difficulty one encounters when
dimensionally regularizing the theory is to give a definition of
the $n\!$-dimensional counterpart of $\ee_{\m\r\n}.$ In this paper we
shall adopt the method proposed by `t Hooft and
Veltman\REF\Hooft{G. `t Hooft and M. Veltman, Nucl. Phys. {\bf B44}
(1972) 189.}[\Hooft] and systematized by Breitenlohner and
Maison\REF\Breitenlohner{P. Breitenlohner and D. Maison, Commun.
Math. Phys. {\bf 52} (1977) 11.}[\Breitenlohner] to incorporate
parity violating objects such as $\ee_{\m\r\n}$
within the framework of dimensional regularization.
This method leads to the only known
algebraically consistent definition of such objects, and has proved
useful in  perturbative computations in WZW\REF\Bos{M. Bos, Ann.
Phys. {\bf 181} (1988)197.}[\Bos] and two-dimensional non-linear
sigma models\REF\Osborn{H. Osborn, Ann. Phys. {\bf 200} (1990)
1.}[\Osborn]. It amounts to defining the $n\!$-dimensional
$\epsilon_{\m\r\n}$ as a completely antisymmetric object in its indices
which satisfies the following identities:
$$
\epsilon_{\m_1\m_2\m_3}\epsilon_{\n_1\n_2\n_3}=
\sum_{\pi\epsilon P_3} {\rm sign}(\pi)\prod_{i=1}^3
{\tilde g}_{\m_i\n_{\pi (i)}} ~~,
$$
$$
\epsilon_{\m_1\m_2\m_3} {\hat g}^{\m_3\m_4}=0 ~~,~~
\epsilon_{\m_1\m_2\m_3} {\hat u}^{\m_3}=0 ~~,
\eqn\definition
$$
$$
u_{\m}={\tilde u}_{\m}\oplus{\hat u}_{\m} ~~,~~
g_{\m\n}={\tilde g}_{\m\n}\oplus{\hat g}_{\m\n} ~~,~~
{\tilde g}^{\m}_{\,\,\,\m}=3~~,~~{\hat g}^{\m}_{\,\,\,\m}=n-3 ~~,
$$
Here all indices run from $1$ to $n,~n$ an intenger larger or
equal than 3, $g_{\m\n}$ is the euclidean metric on $\RR^n,$ and
${\tilde g}_{\m\n}$ and ${\hat g}_{\m\n}$ are its projections onto the
subspaces $\RR^3$ and $\RR^{n-3}.$ For $n\ge 3,$ twiddled and hatted objects
should be understood in this way throughout this paper, \ie\ as projections
onto $\RR^3$ and $\RR^{n-3}.$
Note that this definition of $\ee_{\m\r\n}$ only preserves
$SO(3)\otimes SO(n-3)$ covariance, rather than $SO(n).$
This makes computations more complicated but, on the other hand,
guarantees algebraic consistency. It must be stressed that in dimensional
regularization, when $n$ is promoted to a complex variable $D,$ indices
cease to have particular values and that
$u_\m,~g_{\mn},~\eps_{\m\n\r},\ldots$ are defined through algebraic
relations only [\Breitenlohner], such as \definition.

We can now proceed to dimensionally regularize TMYM theory.
The simplest and most na\"\i ve approach is to promote the 3-dimensional
Feynman rules of TMYM theory to $n$ dimensions and then use dimensional
regularization techniques to evaluate the corresponding integrals.
Despite the simplicity of this prescription, we will show that in
the 3-dimensional limit the BRS identities hold.
There is also the alternative and more sophisticated approach of
constructing an explicitly BRS-invariant regularizing theory.
In what follows we discuss both approaches and show that
they provide the same result.

\medskip

\noindent {\bf 2.1. Na\"\i ve dimensional regularization of TMYM theory}

\smallskip

The idea is to first replace the 3-dimensional Feynman rules
\gaugeprop-\extvert\ with a set of rules in $n$ dimensions.
This is achieved by substituting the Levi-Civita symbol with
the $n\!$-dimensional $\ee_{\m\r\n}$ defined in eq. \definition\
and by considering the 3-vectors and the 3-dimensional euclidean metric
entering in \gaugeprop-\extvert\ as $n\!$-vectors and the
$n\!$-dimensional euclidean
metric on $\RR^n.$ Next, one regards every TMYM diagram as given by
these $n\!$-dimensional Feynman rules, with the proviso that
integrals are formally defined in $n$ dimensions. Finally, one uses the
techniques in [\Breitenlohner ] or \REF\Collins{J. C. Collins,
{\it Renormalization,} Cambridge University
Press (1987).}[\Collins ] to continue Feynman
integrals to complex values $D$ of $n.$ In this way every diagram is
expressed in terms of well-defined integrals for $D$ in some domain
of the complex plane.

The main virtue of this approach is to keep $n\!$-dimensional
propagators as simple as possible so that the algebra is manageable
and, furthermore, the resulting integrals can be computed using
elementary textbook techniques.
All this is very desirable since we will carry out a computation
already cumbersome by itself.
Unfortunately, the dimensionally regularized Green functions
obtained in this way do not satisfy the BRS identities. However, although
BRS invariance is lost at the regularized level, the contributions that
break the BRS identities vanish as $D\to 3.$ Let us make more clear all
these statements.

We begin by obtaining the expression of the
breaking of the BRS identities in terms of Green functions.
It is not difficult to see that the Feynman rules
\gaugeprop-\extvert\ in $n$ dimensions, with $\eps_{\m\n\r}$ as in
\definition, come from the following action:
$$
S[J,H] \,= \,S_m^n[J,H]\, + \, S_B ~~ ,
\eqn\nlaction
$$
where $S_m^n[J,H]$ is local and given by
$$
\eqalign{
S_m^n[J,H]=
& -\,\,i\int d^n\!x ~ \ee^{\m\r\n} \left( {1\over 2}\,\,
                                     A^a_\m \partial_\r A^a_\n +
            {1 \over 3!}\,g\,f^{abc}A^a_\m A^b_\r A^c_\n  \right) \cr
    & + {1\over 4m}\int d^n\!x ~F^a_{\mn}F^{a \,\mn}
           - \int d^n\!x ~s\left(\bar{c}^a\partial A^a
                  +J^aA^a - H^a c^a \right)  \cr
}
\eqn\mnaction
$$
and $S_B$ stands for the non-local contribution
$$
S_B = {1\over 2} \int d^n\! x\,d^n\! y~
                    A^a_\m(x) \, O_{\m\n}(x-y)\, A^a_{\n}(y) ~~.
$$
The operator $O_\mn(x-y)$ is given by
$$
O_{\m\n}(x-y)=\int~{d^n p\over (2\pi)^n}~O_{\m\n}(p)\,
 e^{ip(x-y)}~~,
$$
with
$$
\eqalign{
O_{\m\n}(p)\,=\,{m\over (p^2)^2+m^2\tp^2}\, & \Bigl[\,
       \hp^2 \, \left( -m\ee_{\m\r\n}p^{\r} + p^2 g_{\m\n}
             - p_{\m}p_{\n} \right)  \cropen{10pt}
& + (p^2+m^2)\, \Bigl( {\hp^2\over p^2}p_{\m}p_{\n} + \tp^2{\hat g}_{\mn}
       + \hp_{\m}\hp_{\n} - p_{\m}\hp_{\n}-\hp_{\m}p_{\n} \Bigr) \,
             \Bigr] ~~ . \cr
}
\eqn\Oterm
$$
The BRS operator $s$ in \mnaction\ is the $n\!$-dimensional
version of that in three dimensions: it is given by eqs. \brstrans\ but
now all functions and vectors are defined on $\RR^n.$
In the remainder of this section the operator $s$ should be understood
in this $n\!$-dimensional sense.

To find the BRS identities we need the formal, perturbative
partition function \nextline $Z[j,w,\bar w,\lambda;J,H]$
corresponding to the action $S[J,H]:$
$$
\eqalign{
Z\,[j,w,\bar w,\lambda;J,H]\,=&\,\,{1\over Z[0]} \,
     \int [dA] \,[dc] \,[d{\bar c}]\, [db] ~ e^{-S[J,H] +
          \int d^n\!x \,(jA + \lambda b + {\bar w} c + {\bar c} w)}\cr
= & \sum_{E=0}^{\infty} \int d^{n_1}\!x \cdots d^{n_E}\!x ~
     G(x_1,\ldots,x_E) \,f(x_1)\cdots f(x_E)~~.\cr
}
\eqn\partfunct
$$
Here $f(x)$ denotes any of the sources
$j^a_\m,\,w^a,\,{\bar w}^a,\,\lambda^a$ and/or the external fields
$J^a_\m,\,H^a,$ and the Green functions
$G(x_1,\ldots,x_E)$ are obtained by adding all non-vacuum
Feynman diagrams with appropriate elementary fields and insertions of
type $sA^a_{\m}$ and $sc^a.$ It is immediate to see that the BRS
variation of the local action $S_m^n[J,H]$ vanishes so that
$$
s\,S[J,K] \, = \, s\,S_B = g f^{abc}
    \int d^n\! x \, d^n\! y ~ A^a_\m(x)\, O_\mn(x-y) \,
         A^b_\n(y)\, c^c(y) ~~ .
$$
This, together with the fact that the partition function is invariant
under trnasformations \brstrans, leads to
$$
\Sigma \,Z\,[j,w,\bar w,\lambda;J,H]\, = B ~~,
\eqn\brokenBRS
$$
where $\Sigma$ is the functional operator
$$
\Sigma=\int d^n\! x \,\left( -\,j^a_{\m}\,\,\dj
          + {\bar w}^a\,\dH - w^a\,{\delta\over\delta \lambda^a} \right)
\eqn\sigmaop
$$
and $B$ has the form
$$
B=\,\ {1\over Z[0]} \int [dA] \,[dc] \, [d{\bar c}]\, [db]~(s\,S_B)
   e^{-\,S[J,H]\, + \int d^n\! x \,(jA + \lambda b + {\bar w} c
                        + {\bar c} w)}~~.
\eqn\breaking
$$
Eqs. \brokenBRS\ and \breaking\ are both to be understood as the
perturbation series obtained from the corresponding formal
$n\!$-dimensional Feynman diagrams. That the formal manipulations leading
to eqs. \brokenBRS\ and \breaking\ make sense for the
dimensionally regularized Green functions in \partfunct\ follows
from properties of dimensionally regularized Feynman integrals
[43]. Thus, eqs. \brokenBRS\ and \breaking, when interpreted in terms
of Feynman diagrams, hold in dimensional regularization, where $n$ is
promoted to a complex parameter $D.$

Once the breaking $B$ of the BRS identities has been obtained, the next step
is to show that it goes to zero as $D$ approaches 3. In doing
so one has to prove that the limit $D\to 3$ of the Green function
$$
\eqalign{
gf^{abc}  \Biggr< \int & d^D\! x\, d^D\! y ~ A^a_{\mu}(x)\,O_{\m\n}(x-y)\,
                         A^b_{\n}(y)\, c^c(y) \cr
& \left\{\,\prod_{i=1}^{n_1} A^{a_i}_{\m_i}(x_i)\,
           \prod_{i=1}^{n_2} b^{a_i} (x_i)\,
           \prod_{i=1}^{n_3} c^{a_i}(x_i)\,
           \prod_{i=1}^{n_4} {\bar c}^{a_i}(x_i)\,
           \prod_{i=1}^{n_5} sA^{a_i}_{\m_i}(x_i)\,
           \prod_{i=1}^{n_6} sc^{a_i}(x_i) \right\}  \Biggr> ~~ , \cr
}
\eqn\insertedgreen
$$
vanishes for all values of $n_1,n_2,n_3,n_4,n_5$ and $n_6$ at
any order in perturbation theory, the symbol $<\cdots>$ denoting
$$
{1\over Z[0]}\,\int [dA]\, [dc]\, [d{\bar c}]\, [db] ~\cdots~ e^{-S[0,0]}
{}~~ .
$$
Now, any dimensionally regularized contribution to the the Green function
\insertedgreen\ can be arranged in the form
$$
\eqalign{
I^{a_1\cdots a_E}_{\m_1\cdots\m_A}(p_1,\cdots,p_E) = g f^{abc}
    \int & d^D\! k\,d^D\! q~ D^{a b_1}_{\m\r}(k)\,O_{\r\n}(k)\,
          D^{b b_2}_{\n\s}(q)\,D^{c b_3}(q-k) \cr
& \times G^{b_1 b_2 b_3 a_1\cdots a_E}_{\m\s\m_1\cdots \m_A}
    (k,q,q-k,p_1,\ldots,p_E) ~~ , \cr
}
\eqn\amputated
$$
where $G^{b_1 b_2 b_3 a_1\cdots a_E}_{\m\s\m_1\cdots \m_A}
(k,q,q-k,p_1,\ldots,p_E)$ is some appropriate Green function of the
fields and their derivatives.
The index $A$ denotes the number of external gauge fields,
and $E=\sum_{i=1}^6 n_i$ that of external momenta
$p_1,p_2, \ldots ,p_E.$  Note that $E\ge 2,$ since rigid invariance
under $SU(N)$ and ghost number conservation
imply that both $n_1+n_2$ and $n_3+n_4$ are larger than or equal to $1.$
In what follows we prove that
$I^{a_1\cdots a_E}_{\m_1\cdots\m_A}(p_1,\ldots,p_E)$
goes to zero as $D$ goes to 3 so that $B$ vanishes in this limit.

{}From eq. \amputated\ it follows that every dimensionally regularized
diagram contributing to the Green function \insertedgreen\ contains
the contraction of the gauge propagator $D_{\m\r}^{ab_1}(k)$ with
the operator $O_{\r\n}(k).$ Some trivial algebra yields (dropping
colour indices):
$$
{\hat Y}_{\m\n}(k) \equiv D_{\m\r}(k)\,O^\r_{~~~\n} (k)=
   {m^2\over k^2(k^2+m^2)} \, \left( \hk^2 g_{\m\n} +
        \tk^2 {\hat g}_{\m\n} + \hk_{\m}\hk_{\n} -
           \hk_{\m}k_{\n} - k_{\m}\hk_{\n} \right) ~~ ,
\eqn\simplification
$$
and so the corresponding dimensionally regularized integrals
only involve in their denominators factors of
the type $(l^2)^{r_1}(l^2+m^2)^{r_2},$ with $r_1$ and $r_2$
non-negative integers. The operator ${\hat Y}_{\m\n}(p)$
is evanescent since it goes to zero as $D\rightarrow 3$ [\Collins ].
This does not mean, however, that the integrals
contributing to $I_{\m_1\cdots\m_A},$
all of which contain one $\hat{Y}_{\mn}(k),$ vanish as $D\to 3.$
Indeed, integration over momenta could yield poles at $D=3$ that,
conspired with the evanescent character of ${\hat Y}_{\m\n}(k),$
could give a non-vanishing result. A way to decide whether or not
contributions containing evanescent operators go to zero as $D\to 3$
is to use the following property of dimensional regularization
[\Collins]: if
$$
K_{\m_1\cdots\m_r}(D) \equiv \int d^D\!k ~
    k_{\m_1}\cdots k_{\m_r} ~ f(\tilde{k},\hat{k},p)
$$
is finite \foot{In the case at hand, the function $f$
in the integrand of $K_{\m_1\cdots\m_r}$ only depends on the
$D\!$-dimensional vector $k^\m.$ The property is, however, more general
and englobes the case in which $f$ is a function of $\tilde{k}^\m$
and $\hat{k}^\m,$ which we will meet in Subsect. 2.2.} at $D=3$ then
$$
\lim_{D\to 3} ~ \hat{g}^{\m_1\n_1}\cdots\hat{g}^{\m_r\n_r}\,\,
      K_{\m_1\cdots\m_r}(D)=0 \,\, .\eqn\poles
$$
A sufficient condition for $K_{\m_1\cdots\m_r}$ to be finite at $D=3$
is absolute convergence, which in turn can be established by using
Lowenstein and Zimmerman's power counting
theorem\REF\LZ{J. H. Lowenstein and W. Zimmerman, Comm. Math. Phys.
{\bf 44} (1975) 73.}[\LZ ]. In applying this theorem some care
should be taken and only scaling dimensions should be considered
when counting powers. In other words, one should first express
evanescent objects in terms of $D\!$-dimensional quantities contracted
with $\hat{g}_{\mn},$ {\it e.g.} $\hk^2=\hat{g}^{\m\n}k_\m k_\n,$ and
secondly apply the theorem at $D=3$ prior to contractions.

Property \poles\ and the fact that all terms in ${\hat Y}_{\mn}(k)$
have at least one ${\hat g}_{\mn}$ imply that
to prove that the breaking $B$ of the BRS identities approaches zero
as $D\to 3$ it is enough to show that the integrals
in \amputated\
are finite at $D=3.$ Let us see that this is the case.
Consider a 1PI Feynman
diagram contributing to \amputated, hence to the Green function
\insertedgreen.
Its superficial UV degree $\Omega$ at $D=3$ is given by
$$
\Omega=2-{1\over 2}\,\left[
     n_1+n_2+2(n_3+n_4+n_6)+3n_5+V_3+V_{gh}+2V_4\right] ~~ .
\eqn\nlUVdegree
$$
Since $n_1+n_2\, ,\,n_3+n_4 \ge 1,\,\,\Omega$ satisfies the following
inequality
$$
\Omega\leq {1\over 2}- {1\over 2}\,
     \left(V_3+V_{gh}+2V_4+3n_5+2n_6\right) ~~ .
\eqn\inequality
$$
For one-loop diagrams, the RHS in eq. \inequality\ is in general
positive, so that they are not finite by power counting.
However, Speer \REF\Speer{E. R. Speer, Ann. Inst. Henri
Poincar\'e {\bf XXII} (1975) 1.\subpar E. R. Speer J. Math.
Phys. {\bf 15} (1974) 1.}[\Speer] has shown that in dimensional
regularization no poles arise in one-loop integrations for $D\to n,~(n$
odd) even when the integrals are not convergent by power counting. His
results only apply to integrals whose denominators are products of the
type $(p^2)^{r_1} (p^2+m^2)^{r_2},$ with $r_1$ and $r_2$ non-negative
integers, which is why the simplification \simplification\ has to be
performed beforehand. Thus, one-loop integrals in \amputated\
are free of singularities. As for higher-loop diagrams, they involve,
at least, either two vertices of the type $V_3,$ $V_{gh}$ and $V_4$
or two operators of the form $sA^a_{\m}$ and $sc^a$ $(n_5+n_6 \ge 2)$
so that $\Omega < 0.$ The only possible sources of UV divergences
beyond one loop are then subdiagrams. Now, one-loop subdiagrams are
not a problem since they are finite by Speer's results. The situation is
more complicated for two-loop subdiagrams, since 1PI two-loop diagrams
contributing to the gauge field propagator are not UV finite by power
counting [see discussion following eq. \uvdegree]. However, as
we shall show in Sect. 5, their sum is finite for $D\to 3$ so that
two-loop UV subdivergences cancel. To summarize, 1PI Feynman integrals
in \amputated\ do not give rise to UV poles as $D\rightarrow 3.$
Obviously, the same statement holds for integrals which are not 1PI.

Let us finally see that no IR singularities occur in
$I_{\m_1\cdots\m_A}$ as $D$ goes to $3.$ IR singularities at $D=3$ may
come either from integration over the regions
$$
{}~~ k^2\sim 0\, ,\,q^2>0 ~~ ;
{}~~ k^2>0\, ,\,q^2\sim 0 ~~ ;
{}~~ k^2\sim 0\, ,\, q^2 \sim 0 ~~ ;
{}~~ (k-q)^2\sim 0~~,
\eqn\regions
$$
or from subintegrations in $G^{b_1 b_2 b_3 a_1\cdots a_E}_
{\m\s\m_1\cdots \m_A} (k,q,q-k,p_1,\ldots,p_E).$ Pisarski and Rao
[\Pisarski] have shown that the latter subintegrations do not give
rise to IR singularities. In turn, from eq. \simplification\ one may
readily see that $\hat{Y}^{\m\n}(k) \sim  const$ as $p$ goes to zero, and
so, for IR divergences to take place at $D=3,$ the behaviour of
the Green function $G^{b_1 b_2 b_3 a_1\cdots a_E}_{\m\s\m_1\cdots \m_A}
(k,q,q-k,p_1,\ldots,p_E)$ should be singular in one of the regions
\regions. Now, according to ref. [\Pisarski], this only happens if all
independent momenta in $G^{b_1 b_2 b_3 a_1\cdots a_E}_{\m\s\m_1\cdots \m_A}
(k,q,q-k,p_1,\ldots,p_E)$ go to zero at the same time, but this is not
possible in our case since we assume that the external momenta
$p_1,\ldots,p_E$ are non-exceptional. This completes the proof of
finiteness at $D=3$ of the integrals $I_{\m_1\cdots\m_A},$ which in turn
implies the vanishing of the Green function \insertedgreen, hence of the
breaking $B$ so that BRS invariance is recovered when the limit $D \to 3$
is taken.

\medskip

\noindent{\bf 2.2. BRS-invariant dimensional regularization
of TMYM theory}

\smallskip

One may argue that it is desirable to preserve BRS invariance at the
regularized level, since the BRS symmetry controls gauge invariance
for the gauge-fixed theory. We next propose a method to dimensionally
regularize TMYM theory which explicitly preserves BRS invariance.
The starting point is the action $S_m^n[J,H]$ in eq. \mnaction. The
Feynman rules for this action are the same as in \gaugeprop-\ghostvert,
except for the gauge propagator that now reads
$$
\eqalign{
\dd^{ab}_{\mn}(\tp,\hp) = & {\d^{ab}\, m \over (p^2)^2+m^2\, \tp^2} \,\,
     \Bigl[ \,\,  m\,\ee_{\m\r\n}\,p^\r + p^2 g_{\mn} - p_\m p_\n  \cr
&\qquad\qquad  + {m^2\over p^2}\,\left( {\hp^2 \over p^2} \,\, p_\m p_\n
     +\tp^2 \,\hat{g}_{\mn} - \hp_\m \hp_\n +
           p_\m \hp_\n + \hp_\m p_\n \right)\,\Bigr] ~ . \cr}
\eqn\ndprop
$$
The rest of the method proceeds along the same lines as before. With
this new set of Feynman rules one first constructs $n\!$-dimensional
diagrams. Then, one promotes $n$ to a complex variable $D$ so
as to express Feynman integrals in terms of well-defined functions
for $D$ in some domain of the complex plane. Finally, the value of the
integrals outside the domain is obtained by analytic continuation.
When constructing dimensionally regularized diagrams it must
be taken into account that the denominator $(p^2)^2+m^2\,\tp^2$ makes it
necessary to use dimensional regularization as defined in
refs.$\!\!\!\!\!$\REF\Veltman{G. 't Hooft and M. Veltman, Nucl. Phys.
{\bf B44}(1972) 11.}\REF\Speerb{E. R. Speer, {\it Dimensional and
analytic renormalization}, in {\it Renormalization theory}, G. Velo
and A.S. Wightman eds., Reidel Dordrecht (1976).}[\Collins,\Veltman,\Speerb],
and not as in [\Breitenlohner ].

Let us see that the dimensionally regularized theory is BRS
invariant. On the one hand, $n\!$-dimensional BRS transformations
do not depend on $n$ explicitly and, as already mentioned, leave the
action \mnaction\ invariant.
On the other, since the propagators $\dd^{ab}_{\m\n}(\tp,\hp),$
$D^{ab}_{\m}(p)$ and $D^{ab}(p)$ have been computed by inverting
the kinetic term in the action \mnaction,
the insertion of the operators
$\ee_{\m\r\n}\partial_{\r}A^a_{\n}-\partial_{\m}b^a,$
$\partial^2 c^a$ and $\partial^2 \bar{c}^a$ in a
line of a dimensionally regularized diagram
is equivalent to the contraction of the line to
a point. From these two facts it follows [\Breitenlohner ] that the
dimensionally regularized theory
is consistent with an action principle and that,
therefore, the corresponding Green functions
satisfy the formal BRS identities.

The non-trivial dependence of the gauge field propagator \ndprop\
on $\tp^\m$ and $\hp^\m$ makes calculations
very complicated. However, when the limit $D\to 3$ is taken,
the set of Feynman rules \ndprop\ and \ghostprop-\extvert\
generate the same perturbation expansions as the set \gaugeprop-\extvert.
Here we will show that this is the case at one loop by using
property \poles, and then recall general results from renormalization
theory to extend the result to higher loops.
To state the problem more clearly it is convinitent to write
\ndprop\ as the sum
$$
\dd^{ab}_{\mn}(\tp,\hp) = D^{ab}_{\mn}(p) + R^{ab}_{\mn}(\tp,\hp)  ~ ,
\eqn\mastereq
$$
with $D^{ab}_{\mn}(p)$ as in eq. \gaugeprop\ and
$R^{ab}_{\mn}(\tp,\hp)$ the evanescent operator
$$
\eqalign{
R^{ab}_{\mn}(\tp,\hp) = & {m^3 \over p^2\,[(p^2)^2+m^2\,\tp^2]}
   \,\, \biggl[ \,\, {\hp^2 \over p^2+m^2}\,
         \left( \ee_{\m\r\n}\,p^\r + p^2 g_{\mn} +
                {m^2 \over p^2}\, p_\m p_\n \right) \cr
& \qquad \qquad + \tp^2 \hat{g}_{\mn} + \hp_\m \hp_\n -
              p_\m\hp_\n - \hp_\m p_\n  \,\, \biggr] ~. \cr }
$$
Then, any 1PI Feynman diagram $\F(D)$ with $I_g$ internal gauge lines
can be recast as
$$
\F(D)=\sum_{j=0}^{I_g}\,\F_j(D) ~ , \eqn\eftwo
$$
where $\F_j(D)$ is the sum of all terms with $j$
evanescent operators
$R^{ab}_{\mn}$ and $I_g-j$ factors $D^{ab}_{\mn}.$ In particular,
$\F_0(D)$ is obtained from $\F(D)$ by replacing the full propagator
\ndprop\ with the simpler \gaugeprop\ in all internal gauge lines.
Now, any 1PI Green function ${\cG}(D)$ is obtained by summing over all
topologically non-equivalent 1PI Feynman diagrams with appropriate
symmetry factors. Thus, ${\cG}(D)$ can be written as
$$
{\cG}(D)\,=\, {\cG}_0 (D)\,+\,\hat{\cG}(D) ~,
$$
with ${\cG}_0 (D)$ the Green function as obtained from
the Feynman rules \gaugeprop-\extvert.
The aim is to show that $\hat{\cG}(D)$ goes to zero as
$D$ appraoches 3.

We begin by using property \poles\ to show that at one loop,
$\F_j(D) \to 0$ as $D\to 3$ for $j\ge 1,$ which in turn implies
that $\hat{\cG}(D)$ goes to zero as $D$ approaches 3.
Property \poles\ requires finiteness at $D=3$ so
a detailed analysis of the convergence properties of
$\F_j(D)$ is in order.
The UV degree of $\dd^{ab}_{\mn}(\tp,\hp)$ is the same as
that of $D^{ab}_{\mn}(p),$ namely $-2,$  ---the degree used
to compute $\omega_m$ in eq. \uvdegree.
In turn, $R^{ab}_{\mn}(\tp,\hp)$ has UV degree $-4,$
\ie\ two units less, and so the overall UV degree of $\F_j(D)$ at
$D=3$ is $\omega_m -2j.$ Diagrams with only one external gauge
leg vanish because of colour algebra (see Sect. 4). Also,
there are no diagrams with only one external ghost leg, since they
would violate ghost conservation number. Using then
that diagrams with two or more external legs have $\omega_m \leq 1$
we conclude that $\F_j(D)\,\, (j\ge 1)$ is overall UV convergent
at $D=3.$ As for IR convergence, the IR degrees of
$\dd^{ab}_{\mn}(\tp,\hp),$ $D^{ab}_{\mn}(p)$ and
$R^{ab}_{\mn}(\tp,\hp)$ are respectively $-2,\,\,-1$ and $-2.$
The IR behaviour of $\F_j(D)$ for all $j$ is then at least
as good as that of $\F(D),$ and thus if the full diagram $\F(D)$
is overall IR finite by power counting so are all the terms $\F_j(D).$
Putting everything together, we have that if a one-loop diagram
does not have IR divergences, property \poles\
implies that $\lim_{D\to 3}\F_j(D)=0$ for $j\ge 1.$ Now, power counting
shows that one-loop diagrams are IR convergent at $D=3$ for
non-exceptional external momenta, something very easy to check for
the diagrams we are going to be concerned with (see. Figs. 2, 7 and 9).
It then follows that for one-loop diagrams, $\F_{j\ge 1}(D)$ goes to
zero as $Dto 3$ and that we can use \gaugeprop\ as the gauge
field propagator to obtain their non-vanishing contributions in this limit.
Therefore, as $D$ approaches 3, one-loop diagrams only
involve denominators of the type
$(p^2)^{r_1}(p^2+m^2)^{r_2},$ with $r_1$ and $r_2$
non-negative integers, Speer's [\Speer] conclusions apply and,
as a result, one-loop diagrams are finite at $D=3.$

To show that $\hat{\cG}(D)$ vanishes as $D\to 3$ at the two-loop order
we shall take advantage of two general results from perturbative
renormalization\REF\Hepp{K. Hepp, {\it Renormalization theory} in
{\it Statistical mechanics and quantum field theory,} Les Houches 1970,
C. DeWitt and R. Stora eds., Gordon and Breach (1971).\subpar
H. Epstein and V. Glaser, Ann. Inst. Henri Poincar\'e {\bf XIX} 3
(1973) 211.\subpar
D. Maison, {\it Renormalization theory, a short account of results and
problems} in {\it Renormalization of quantum field theories with
non-linear field transformations}, P. Breitenlohner, D. Maison and
K. Sibold eds., Springer Verlag (1987).}[\Collins,\Hepp]. Let us
consider a 1PI Green function that is UV divergent at the $(n+1)\!$-loop
order and that has been renormalized up to order $n.$ The first result
states that the UV divergent part of the function at order
$n+1$ is a local polynomial in the external momenta, with degree
at most equal to the superficial UV degree of the function at $n+1$ loops.
Let us state the second result. We shall consider the renormalized
values of a 1PI Green function obtained by using two different
renormalization schemes and assume that these values are equal up to
the $n\!$-loop order. Then, their difference at order $n+1$
is a polynomial in the external momenta if the superficial UV
degree of the function at $n+1$ loops is equal or larger than zero.
The degree of this polynomial is at most equal to the UV degree of
the function at $n+1$ loops. If the superficial UV degree of the
function at $n+1$ loops is negative, then the difference vanishes.

We have seen that both dimensional regularization
methods introduced in this and the previous subsections
yield a finite limit as $D\rightarrow 3$ at one loop.
These limits define therefore two renormalization schemes at
the one-loop order. Moreover, these two renormalization
schemes agree at this order, since we have seen that both $D\to 3$ limits
are equal at one loop. Hence, the two results above ensure that
the limit $D\rightarrow 3$ of any 1PI function, but for the
vacuum polarization tensor, is finite at the two-loop
level and that it is the same for both dimensional regularization methods,
\ie\ $\hat{\cG}(D)\rightarrow 0$ as $D\rightarrow 3.$

The two-loop vacuum polarization tensor is logarithmically overall UV
divergent. Thus, the difference between the vacuum polarization tensor
in the regularization of Subsect. 2.1, $\Pi^{(2.1)}_{\m\n},$ and the
vacuum polarization tensor for the regularization introduced in the
present subsection, $\Pi^{(2.2)}_{\m\n},$ is
$$
\Pi^{(2.1)}_{\m\n}\,-\,\Pi^{(2.2)}_{\m\n}\,=\,
m\,(a_1\, {\tilde g}_{\m\n}\,+\,a_2\, {\hat g}_{\m\n})\,+\, O(D-3) ~ .
\eqn\difference
$$
A number of comments regarding eq. \difference\ are in order. First, the
overall factor $m$ occurs since the mass dimension of the vacuum
polarization tensor is one, despite its superficial UV degree is zero.
Secondly, the fact that the coefficients $a_1$ and $a_2$ might not be
equal is a consequence of the fact that our definition for the
$n\!$-dimensional $\ee_{\m\r\n}$ renders the formal theory invariant
under $SO(3)\otimes SO(n-3)$ rather than $SO(n)$ [\Bos].
Finally, we have not made any hypothesis on the finiteness
of $a_1$ and $a_2$ as $D\rightarrow 3.$  Now, we have seen that both
dimensional regularization methods yield BRS-invariant 1PI functions
as $D\rightarrow 3.$ Hence, in the limit $D\rightarrow 3$ we have:
$$
p^{\m}\,\Pi^{(2.1)}_{\m\n}\,=p^{\m}\,\Pi^{(2.2)}_{\m\n}\,=\, 0\,\,.
\eqn\transversality
$$
Eqs. \difference\ and  \transversality\ lead to
$$
         a_1\,=\,a_2\,=0\,\,,
$$
and so, $\hat{\cG}(D)$ goes to zero as $D$ approaches 3
for the two-loop vacuum polarization tensor. We have then shown that
the two-loop vacuum polarization tensor computed with any of the
regularization methods in this section is the same, modulo
contributions that vanish in the limit $D\to 3.$ It is worth
noticing that the general arguments we have used to show that
at two loops $\Pi^{(2.1)}_{\m\n}\,-\,\Pi^{(2.2)}_{\m\n}$ goes to zero as
$D\rightarrow 3$ would have left behind a non-vanishing
contribution of the type
$$
a_3\,\ee_{\m\r\n}\,p^{\r}\,\, ,
$$
had we applied them at the one-loop level.

Let us analyze now what happens beyond the two-loop order. Now the
superficial UV degree of any 1PI function of TMYM theory is negative.
We shall see in Sect. 5 that the limit $D\rightarrow 3$ of the two-loop
vacuum polarization tensor is actually finite. Then, putting
everything together, we conclude that the limit $D\rightarrow 3$ of
any 1PI function of TMYM theory is finite at $n$ loops. Moreover, the
latter limit is the same for both regularization methods in this section.
This shows UV finiteness for TMYM theory at any order in the loop
expansion (see also Sect 7).

To summarize, only the non-evanescent part \gaugeprop\ of the
gauge field propagator \ndprop\ contributes to the limit $D\to 3$
of 1PI Green functions, the objects we will compute here.
It is important to emphasize, however, that the starting point is the
full propagator \ndprop, which ensures BRS invariance at the
regularized level, and that it happens a posteriori that as $D$ goes to 3
only its non-evanescent part contributes.

Notice that for any of the two dimensional regularizations we have defined,
the 3-dimensional limit of dimensionally regularized TMYM theory is
BRS-invaraint. Once this has been established, it follows that the limit
$m\to\infty$ preserves BRS invariance, since the regulator $m$ does not
enter explicitly in the BRS identities.

Here we adopt the BRS-invariant definition of dimensional regularization
of this subsection. In the remainder of the paper, dimensionally
regularized quantities should be understood to be defined in
this sense. As far as explicit computations are concerned, both
approaches give the same result. However, from the point of view
of understanding the results, they provide different pictures,
as will become clear in Sect. 5.

In this paper we are going to evaluate the limit $D\to 3,\,\, m\to\infty$
of some dimensionally regularized TMYM Green functions.
As is by now clear, their 3-dimensional limit only receives
contributions from non-evanescent objects so that to compute the
relevant Feynman diagrams we will use \gaugeprop\ for the gauge field
propagator. As for the large $m$ limit, we present
in the next section two vanishing theorems that will be very helpful
in computing it.

{\bf \chapter{Large mass vanishing theorems}}

In this section we present two power counting-like theorems which
provide sufficient conditions  for an integral arising from a Feynman
diagram to vanish in the limit $m\to\infty.$ These theorems will
be widely used in Sects. 5 and 6 below. They will enable us not only
to disregard integrals that do not contribute to the limit
$m\to \infty$ without having to compute them, but also to
explicitly evaluate non-zero contributions.

In what follows we will be concerned with integrals of the form
$$ I(p,m) = \,m^{\beta}\,\int dk \,\, F(k,p,m) \,\, , \eqn\ipm $$
where the integration measure is
$$ dk = d^N\!k_1\cdots d^N\!k_M $$
and the integrand is a rational function:
$$
F(k,p,m) \, = \, {{M(k)} \over
          {\prod_{i\in H} (l_i^2 +m_i^2)^{n_i}}}  ~ .
\eqn\integrand
$$
The greek letter $\beta$ denotes an arbitrary real number and $N$ and $n_i$
are positive integers. The numerator $M(k)$ is a monomial of degree $n_k$
in the components of the vectors $k_1,\ldots , k_M.$ The $l_i$'s are
linear combinations
$$ l_i = K_i(k) + P_i(p) \, \, , $$
$$ K_i(k)= \sum_{j=1}^M a_{ij} k_j \quad , \quad
       P_i(p)= \sum_{j=1}^E b_{ij} p_j \,\, , $$
with not all the coefficients $a_{ij}$ vanishing for any given $i.$
We will assume that the external momenta
$p_1,\ldots ,p_E$ lie
in a bounded subdomain of ${\RR}^{D}$ and consider
the case in which the masses only
take on two values, namely $m_i=0$ for, say, $i\in S_0$ and
$m_i=m>0$ for $i\in S_1=H\setminus S_0.$ We want to find
sufficient conditions for the integral \ipm\ to go to zero as
$m\to\infty$ so that we can assume without loss of generality that $m/p>1.$

We call $d$ to the mass dimension of $I(p,m)$ and denote by
$\ird_{\,\min}$ the minimum of zero and the infrared degrees of all
the subintegrals of $I(p,m)$ at zero external momenta, including
$I(p,m)$ itself. Then the following theorem holds:

{
\leftskip=1 true cm \rightskip=1 true cm
\noindent {\bf $m\!$-Theorem:} {\sl If the  integral $I(p,m)$ is both
UV and IR covergent by power counting at non-exceptional external momenta,
and the mass dimension $d$ and $\ird_{\,\min}$ defined above satisfy
$$
d - \ird_{\,\min} \, < \, 0  ~ ,
\eqn\mth
$$
then $I(p,m)$ vanishes when $m$ goes to $\infty.$}
\par}

\noindent
To prove the theorem we will proceed in two steps. We will
first bound $I(p,m)$ from above keeping $m/p>1$ fixed, the bound in
general depending on $m,$ and then find conditions for the bound to
approach zero as $m\to\infty.$

Following Lowenstein and Zimmermann [\LZ] we write the
integral \ipm\ as
$$ I(p,m) = \, \sum_S I_S(p,m) \, \, ,$$
where
$$ I_S(p,m) = \,m^{\beta}\,\int_{\scriptstyle l_i^2<m^2\,\,\,\,i\in S
              \atop \scriptstyle l_i^2\geq m^2\,\,\,\, i\in T}
              \!\! dk\,\, F(k,p,m)  $$
and the sum is extended to all subsets $S$ of $S_0.$ $T$ is the
complement of $S$ in $S_0,$ \ie\ $T=S_0 \setminus S.$ To study
$I_S(p,m),$ we note that among the $l_i\!$'s occuring in the denominator
of \integrand\ we can choose without loss of generality $M$
of them such that the change of variables
$$ (k_1,\ldots , k_M) \longrightarrow
  \left( u_1=l_{i_{\scriptstyle 1}},\ldots ,
        u_a=l_{i_{\scriptstyle a}}, v_1=l_{i_{\scriptstyle a+1}},
        \ldots , v_b=l_{i_M}\right) $$
has jacobian one. Furthermore, $l_{i_1},\ldots ,l_{i_M}$ can be chosen
in such a way that the homogeneous parts in $k_i$ of
$u_1,\ldots ,u_a$ form a basis of the homogeneous parts in $k_i$ of
the $l_j\!$'s, $j\in S.$ In terms of the new variables, the numerator
$M(k)$ in \integrand\ reads
$$M(k) = \sum_{\alpha} A_{S\alpha}(v,p)\,\,M_{\alpha}(u) \,\, ,$$
with $M_{\alpha}(u)$ a monomial in the components of
$u_1,\ldots , u_a$ and $A_{S\alpha}(v,p)$ a polynomial in those of
$v_1,\ldots , v_b$ and the external momenta $p_1,\ldots , p_E.$
As for the vectors $l_i,$ we have that
$$ l_i=U_i(u)+Q_i(p) \qquad  {\rm if} \quad i\in S\,\, ,$$
$$ l_i=V_i(v)+R_i(u,p)\qquad  {\rm if} \quad i\not\in S\,\, ,$$
with $U_i,\>Q_i,\>V_i$ and $R_i$ linear combinations of their arguments.
For the integral $I_S(p,m)$ we obtain
$$ I_S(p,m) =\,m^{\beta}\,
 \sum_{\alpha}
        \int_{\scriptstyle l_i^2<m^2\,\,\,i\in S}\!\!
           du\,\,\, {{M_{\alpha}(u)} \over {\prod_S(l_i^2)^{n_i}}}
        \int_{\scriptstyle l_i^2\geq m^2\,\,\,i\in T}\!\!
           dv\,\,\, {{A_{S\alpha}(v,p)} \over {\prod_T(l_i^2)^{n_i}\,
         \prod_{S_1} (l_i^2+m^2)^{n_i}}}\,\, . $$
Note that when integrating over $v,$ the $R_i\!$'s play the r\^ole of
bounded external momenta, since the vectors $u_i$ and $p_i$ are
themselves bounded. From this observation, the fact
that $p/m<1$ and proceeding in the same way as in [\LZ] we get that
$$I_S(p,m) \leq \sum_\a C_{S\alpha}\,\,
           m^{\textstyle d - d_{S\alpha}}
                      \,\,I_{S\alpha}(p,m)\,\, ,$$
where $\{ C_{S\alpha} \}$ are constants that do not depend on $m,$
and $d$ and $d_{S\alpha}$ are
the mass dimensions of the integrals $I(p,m)$ and
$$
I_{S\alpha}(p,m) = \int_{\scriptstyle l_i^2<m^2\,\,\,i\in S}
     \!\! du\,\,\, {{|M_{\alpha}(u)|} \over
             {\prod_S(l_i^2)^{n_i}}}\,\, .
$$
We thus have that
$$
I(p,m) \,\leq\,  \sum_{S,\alpha}\,\, C_{S\a}\,\,
       m^{\textstyle d - d_{S\alpha}}
         \,\, I_{S\alpha}(p,m)\,\, .
$$
Note that $I_{S\alpha}(p,m)$ depends on $m$ through the domain of
integration. Moreover, $m$ is an UV cut-off for the integration variables
$u_i$ so that the integration domain is bounded.
The integrals $I_{S\alpha}(p,m)$ being then covergent
at non-exceptional external momenta. To estimate the large $m$ behaviour
of $I(p,m)$ we have thus been led to evaluate the behaviour
of the massless integrals $I_{S\alpha}(p,m)$ as their UV cut-off $m$
goes to infinity. The latter is usually done
by using Weinberg's theorem\REF\Weinberg{S. Weinberg,
Phys. Rev. {\bf 188} (1960) 838.}[\Weinberg], which establishes that
the leading contribution to $I_{S\alpha}(p,m)$ as $m\to \infty$ is
bounded from above by
$$
C'_{S\alpha}\,\,m^{{\textstyle \uvd^{\,{\rm max}}_{S\alpha}}} ~
\left[ \>{\rm ln}(m/p)\> \right]^{L_{S\alpha}} ~ .
\eqn\bound
$$
Here $C'_{S\alpha}$ is a positive constant, $L_{S\alpha}$ is a
natural number and $d^{\,{\rm max}}_{S\alpha}$ is given by
$$
\uvd^{\,{\rm max}}_{S\alpha} = {\rm max}\,\{0, \{\uvd_i\} \} ~ ,
$$
where the $\uvd_i\!$'s are the UV degrees of all the subintegrals
of $I_{S\alpha}(p,m)$ including itself. Notice that if all the
$\uvd_i\!$'s are negative, the integral $I_{S\a}(p,m)$ is convergent
and the bound \bound\ holds with $L_{S\a}$ zero.

Let us mention that the fact that Weinberg's theorem can be applied to
determine the large $m$ behaviour of $I_{S\alpha}(p,m)$ involves some
technical subtleties. Indeed, the integrals we are studying are massless,
whereas Weinberg considered massive integrals so that his
techniques did not fall short of mathematical rigour. For this reason we
have rigourously checked that if $I(p,m)$ in eq. \ipm\ is either a one- or
a two-loop integral, the large $m$ leading contribution from any of its
$I_{S\alpha}(p,m)$ is indeed bounded from above by \bound. We did so by
using the methods in [\LZ], suitable for massless integrals. To
avoid an unnecessarily long paper, we will not present the details here.
These techniques can be applied to any arbitrary integral of the
type $I_{S\alpha}(p,m).$
We  thus strongly believe that the bound provided by eq. \bound\
holds whatever the number of loops. Let us also stress that in
this paper we are only concerned with one- and two-loop integrals.
After these remarks, we come to the following result:
$$
\lim_{m\rightarrow\infty}
I(p,m) \,\leq\,\lim_{m\rightarrow\infty}
  \sum_{S,\alpha}\,\, C''_{S\a}\,\,
     m^{({\textstyle d - d_{S\alpha} + \uvd^{\,{\rm max}}_{S\alpha}})}
         \,\, \left[\>{\rm ln}(m/p)\>\right]^{L_{S\alpha}} ~ .
$$
Next we note that the UV degree of the integrals $I_{S\alpha}(p,m)$ is
equal to minus their IR degree at zero external momenta, these integrals
being massless. It then follows that
$$
{\textstyle  d_{S\alpha} - \uvd^{\, {\rm max}}_{S\alpha}}\, = \,
{\min}\{0,\ird^{\, \min}_{\,S\alpha}\}\,\, ,
$$
where $\ird^{\, \min}_{\,S\alpha}$ is the minimum of the IR degrees
at zero momenta of $I_{S\alpha}$ and all its subintegrals. Finally,
$$
{\rm min}\,\{{\textstyle  d_{S\alpha} - \uvd^{{\rm max}}_{S\alpha}}\}\,
=\, {\rm min}\,\{0, \{\ird^{\, \min}_{\,S\alpha}\} \}\,=\,
{\ird_{\,\min}}\,\, ,
$$
where the minimum is computed over the set of all $(S,\alpha).$
Putting everything together, we obtain
$$
\lim_{m\rightarrow\infty}I(p,m)\,\leq\,\lim_{m\rightarrow\infty}
\left[ C'''\,\, m^{({\textstyle d - \ird_{\,\min}})}\,\,
\left[ \>{\rm ln}(m/p)\> \right] ^{\,{\rm max}\,\{L_{S\alpha}\}}\right] ~ ,
$$
so that
$$ \lim_{m\rightarrow\infty} I(p,m)\,=\,0 $$
if $d - \ird_{\,\min}\,<\,0.$  This completes the proof of the
theorem.

To formulate the second theorem announced at the beginning
of the section we remind that $n_k$
denotes the degree in $k$ of the numerator in $I(p,m)$
and introduce the symbol
$$
[n]=\left\{ \eqalign{ 1 \qquad &{\rm for} \,\,\, n \,\,\, {\rm odd}\, ,\cr
              0 \qquad &{\rm for} \,\,\, n \,\,\, {\rm even}\, . \cr}
    \right.
$$
The theorem  states the following:

{
\leftskip=1true cm \rightskip=1 true cm
\noindent {\bf $o\!$-Theorem:} {\sl If the  integral $I(p,m)$ in \ipm\
is absolutely convergent at zero external momenta
and its mass dimension $d$ satisfies
$$
[n_k] > d \>, \eqn\oth
$$
then $I(p,m)\to 0$ as $m\to \infty.$}
\par}

\noindent The hypothesis about convergence is more restrictive if
compared to that of the $m\!$-theorem, since now IR convergence for
zero external momenta is required.
Nevertheless, the theorem covers cases which escape the $m\!$-theorem
and in this sense is complementary to the latter.

As for the proof, let us start
by rescaling the integrated momenta $k\to km.$ Eq. \ipm\ thus becomes
$$
I(p,m) = m^{d} \,\,J(p/m) \,\, ,
$$
where
$$ J(p/m) =\int dk \, F(k,p/m,1) \,\, . $$
Since the limit $m\to\infty$ of $J(p/m)$ is well defined by the
hypothesis of abolute convergence at zero external momenta, we
have that
$$
\lim_{m\to\infty}\,\, I(p,m) = J(0) \,\, \lim_{m\to\infty}\, m^{d} ~ .
\eqn\limit
$$
Now, condition \oth\ implies that $d \leq 0$ so that the only
case for which the RHS
in \limit\ is not zero trivially is $d=0,$ which in turn corresponds to
$n_k$ odd. But for $n_k$ odd, $J(0)$ is zero by $SO(N)$ covariance.
This closes the proof of the theorem.

We finish this section with a remark about the hypothesis of UV
convergence in the $m\!$-theorem as applied to TMYM theory.
Let us recall that
we are seeking for results that allow us to establish whether a
dimensionally regularized integral vanish in the limit
$D\to 3,\,\,m\to \infty.$ The most general dimensionally regularized
integral that we will find in the next sections is of the form
$$
I(p,m)= m^{\beta}\,\int d^D\!k\,\,d^D\!q\,\,F(k,q,p,m) \,\, ,
\eqn\intspur
$$
where $k$ and $q$ are the internal momenta and $p$ stands, as usual,
for the external momenta $p_1,\ldots ,p_E.$ The integrand is
$$
F(k,q,p,m)={k^{n_k}\, q^{n_q} \over
    \prod_A [K_i^2\, (K_i^2+m_i^2)]^{n_i} \,\,
       \prod_B [Q_i^2\, (Q_i^2+m_i^2)]^{n_i} \,\,
          \prod_C [R_i^2\, (R_i^2+m_i^2)]^{n_i} } \,\, ,
$$
with $K_i$ a linear combination of $k$ and $p,\,\,Q_i$ of
$q$ and $p,$ and $R_i$ of $k,q$ and $p.$ We shall assume in the
sequel that $I(p,m)$ is IR convergent by power counting
at non-exceptional external momenta and $D=3,$ as is the case for
the two-loop integrals we will find. Regarding UV convergence,
it demands overall convergence and convergence along the $k\!$- and
$q\!$-loops, or more precisely:
$$
\eqalignno{
{\overline\omega} &\equiv 6+ n_k+n_q-4\,\sum_{A,B,C} n_i \, <0 \,\, ,
                                & \eqnalign{\spurall} \cr
&{\overline\omega}_k  \equiv 3+n_k-4\,\sum_{A,C} n_i \, <\, 0 \,\, ,
                                & \eqnalign{\spurk} \cr
&{\overline\omega}_q  \equiv 3+n_q-4\,\sum_{B,C} n_i \, <\, 0 \,\, ,
                                & \eqnalign{\spurq} \cr
}
$$
where we have taken $D=3.$ These conditions are
not always met in our case, since power counting as applied to
TMYM theory yields the following two situations:
$$
\eqalign{
&\bar\omega < 0~~,~~
{\bar\omega}_k = 0,1~~,~~ {\bar\omega}_q < 0\,\, ,\cr
&\bar\omega < 0~~,~~
{\bar\omega}_k < 0~~,~~ {\bar\omega}_q = 0,1\,\, . \cr
}
$$
In both scenarios the integral is overall UV convergent by power counting
but not along one of its subintegrals. In our TMYM computations,
${\overline\omega}_k$ and ${\overline\omega}_q$ are never larger than one.
Notice that if ${\overline\omega} < 0,$ then both ${\overline \omega}_k$
and ${\overline \omega}_q$ cannot be  equal or larger than zero at
the same time. Now, we know that in dimensional regularization one-loop
integrals of the type we are considering are free of singularities
as $D$ goes to 3
[\Speer]. This makes us suspect that conditions \spurk\ and \spurq\
requiring absence of UV subdivergences could be dropped. Let us
show that this is so. Here we present the proof for the case in which
${\overline\omega}_k=0,1.$

To be more specific, we want to prove that as far as the
requirement of UV convergence is concerned, it is enough to
demand \spurall, even when ${\overline\omega}_k=0,1,$
to rightly apply the $m\!$-theorem in the finite limit
$D\rightarrow 3.$ To do this we write $F(k,q,p,m)$ as
$$
F(k,q,p,m)=E(k,q,p,m)\,G(q,p,m)\,\, ,
$$
$E$ and $G$ being given by
$$
E(k,q,p,m)= { k^{n_k} \over
       \prod_A [K_i^2\, (K_i^2+m_i^2)]^{n_i} \,\,
              \prod_C [R_i^2\, (R_i^2+m_i^2)]^{n_i} }
$$
and
$$
G(q,p,m) = { \, q^{n_q} \over
          \prod_B [Q_i^2\,(Q_i^2+m_i^2)]^{n_i} } \,\, .
$$
We further decompose $E$ as
$$
E(k,q,p,m) = E_{\rm fin}^{(l)}(k,q,p,m)
                 + E_{\rm div}^{(l)}(k,m) \,\, ,
\eqn\decom
$$
where $l= {\overline\omega}_k$ and
$$
E_{\rm fin}^{(l)}(k,q,p,m)=\left[ 1 - t^{(l)}(q=0\, ,\,p=0)\right]\,
                    E^{(l)}(k,q,p,m)\,\, ,
$$
$$
E_{\rm div}^{(l)}(k,q,m)=t^{(l)}(q=0\, ,\,p=0)\, E^{(l)}(k,q,p,m)\,\, .
$$
The operators $t^{(l)}$ read:
$$
\eqalign{
t^{(0)}(q=0,p=0)\,H(k,q,p) & =H(k,0,0) \,\, ,\crr
t^{(1)}(q=0,p=0)\,H(k,q,p) & =H(k,0,0) + q^{\m}
      {\left[{\partial H(k,q,p)}\over{\partial q^{\m}}\right]}_{q=p=0}
             + p^{\m}
      {\left[{\partial H(k,q,p)}\over{\partial p^{\m}}\right]}_{q=p=0}
                                     . \cr
}
$$
Note that by construction $E_{\rm fin}^{(l)}$ carries the UV finite part and
$E_{\rm div}^{(l)}$ the divergent (or singular) one. As a matter of fact,
$E^{(l)}_{\rm fin}$ is a BPHZ
subtraction at zero momentum  $q=0\,,\,p=0$\REF\bphz{J.H.
Lowenstein, {\it BPHZ renormalization} in {\it Renormalization
theory}, G. Velo and A. S. Wightman eds., D. Reidel Publishing
Company (1976).}[\bphz].
The decomposition \decom\ gives for $I(p,m):$
$$
I(p,m)=I_{\rm fin}^{(l)}(p,m)+I_{\rm div}^{(l)}(p,m) \,\, ,
$$
with
$$
I_{\rm fin} ^{(l)}(p,m)= \,m^\b\int d^D\!k\,d^D\!q\,\,
     E_{\rm fin}^{(l)}(k,q,p,m)\,\,G(q,p,m) \,\, ,
$$
$$
I_{\rm div}^{(l)}(p,m)= \,m^\b \int d^D\!k\,d^D\!q\,\,
     E_{\rm div}^{(l)}(k,q,m)\,\,G(q,p,m) \,\, .
$$
The integral $I_{\rm fin}^{(l)}(p,m)$ is already UV convergent by power
counting at $D=3.$ Regarding its IR convergence,
we have to make sure that the
zero-momentum subtractions we have performed do not introduce any
IR divergences. Since the subtractions are at zero momenta, we have to
analyze the power counting concerning integration over the regions
$k\sim 0,\,\,q\sim 0,$ and $k\sim 0\, ,\, q\sim 0.$ The IR degree
${\underline\omega}_{\, k}$ of $I_{\rm fin}^{(l)}(p,m)$ in the
region $k\sim 0$ satisfies
$$
{\underline\omega}_{\, k}\,\geq\, 3+n_k-2\,\sum_{A,C} n_i - l\,\, .
$$
Taking into account that
$l={\overline\omega}_k = 3+n_k-4\,\sum_{A,C}\, n_i,$ we conclude that
$$
{\underline\omega}_{\, k} \,\geq\, 2\,\sum_{A,C} n_i \,>\, 0 \eqn\irk
{}~ .
$$
Thus, IR finiteness holds by power counting for the $k\!$-integration in
$I_{\rm fin}^{(l)}(p,m)$ at non-exceptional external momenta and $D=3.$ IR
finiteness at non-expectional momenta for any other domain of
$I_{\rm fin}^{(l)}(p,m)$ holds as well, since $I(p,m)$ is IR finite
by power counting at non-exceptional momenta and $D=3.$ This shows
that $I_{\rm fin}^{(l)}(p,m)$ is both UV and IR convergent by power counting
at non-exceptional momenta and $D=3,$ which in turn implies
that the limit $D\rightarrow 3$ can be taken inside the integral.
As for the other conditions required by the $m\!$-theorem
for $I_{\rm fin}^{(l)}(p,m)$ to vanish as $m$ goes to $\infty,$ it is easy
to see that they follow from those for $I(p,m).$ Concerning
$I_{\rm div}^{(l)}(p,m),$ some simple algebra shows that it is a linear
combination of products of two one-loop integrals. The integrals over $k$
are independent of $p$ and their dependence on $m$ can be scaled away
by rescaling the integration variable $k\to km.$ This confines the initial
UV divergence to integrals independent of $m$ and $p$ which in the limit
$D\to 3$ will produce finite constants ${\cal K}_\a$ [\Speer]:
$$
I_{\rm div}^{(l)}(p,m)= \sum_{\a=0}^l \,\,\,{\cal K}_\a \,\,
      \int d^3\!q\,\, { m^{2\a}\, q^{\a+n_q} \over
            \prod_B [Q_i^2  (Q_i^2+m_i^2)]^{n_i}}
\,\, , \eqn\dimreg
$$
$$
{\cal K}_\a =\lim_{D\to 3}
\int d^D\!k \,\, k^{n_k+\a}\,\,f_{\a}(k^2)\,\, ,
$$
with $f_{\a}(k^2)$ known functions of $k^2.$
Integrals over $q$ in \dimreg\ are UV convergent, since
their UV degree is $3+\a+n_q-2\,\sum_B n_i =\a-l+\overline\omega$ and
$\bar\omega<0$ by hypothesis. They are also IR convergent at non-exceptional
external momenta by hypothesis. Thus we can use the $m\!$-theorem to study
their large $m$ limit. Again, it is a matter of algebra to show
that the conditions that the $m\!$-theorem demands for \dimreg\ to vanish as
$m\to\infty$ follow from those for $I(p,m),$ with the exception
of $\overline\omega_k=l<0,$ which now is no longer needed.

All in all, we have that the integral \intspur\ approaches zero as $m$
goes to $\infty$ if all the conditions demanded by the $m\!$-theorem, except
UV convergence along $k,$ are met. It is straightforward to see that
the same methods show that UV covergence along $k$ can also be dropped
for the $o\!$-theorem.

{\bf \chapter{The one-point Green function for the gauge field}}

Let us consider the gauge field one-point Green function
$<\!A^a_{\mu}(x)\!>$ for TMYM theory with action \maction.
According to eq. \uvdegree, the  integrals that occur
in 1PI Feynman diagrams with only one external gauge line and
no external ghost lines have the highest superficial UV degree.
In fact, one has to go to order $g^6$ to get superficial UV
convergence by power counting. Dimensional regularization, as defined
in any of the two forms in Sect. 2, renders these divergent
diagrams finite. Likewise, IR divergences at non-exceptional
external momenta are regularized. It thus makes sense to
speak of the corresponding dimensionally regularized diagrams.

Every single dimesionally regularized 1PI Feynman diagram that occurs
in the computation of $<\!A^a_{\mu}(x)\!>$ is zero due to colour algebra.
Indeed, its colour factor can always be recast as a linear combination
of terms of the form
$$
{\rm Tr}\,(t^a t^{a_1}\cdots t^{a_i})\,\,
{\rm Tr}\,(t^{a_{i+1}}\cdots t^{a_j})\,\,\cdots
{\rm Tr}\,(t^{a_k}\cdots t^{a_{2N}}) = C^a~,
$$
where all colour
indices $a_1\cdots a_{2N}$ are contracted in pairs and the $t^a\!$
denote the generators in the adjoint representation of $SU(N).$ Obviously,
$C^a = 0.$ This implies that ({\it i}) diagrams with gauge field one-point
subdiagrams are zero and that ({\it ii}) the Green function
$<\!A^a_{\mu}(x)\!>$ vanishes. The latter is the same as to say that
there are no contributions to the TMYM regularized effective action with
only one gauge-field external leg. By taking then into account the definition
in eq. \mlimit\ one concludes that the effective action for perturbative
CS field theory has no one-point term for the gauge field.

{\bf\chapter{The gauge field vacuum polarization tensor}}

The purpose of this section is to describe how we have carried out the
computation of the vacuum polarization tensor
$\Pi^{ab}_{\mn}(p)$ for $SU(N)$
CS theory in the Landau gauge at second order in
perturbation theory. This entails as a first step the calculation of the
limit $D\rightarrow 3$ of the dimensionally regularized topologically
massive $SU(N)$ two-loop vacuum polarization tensor
$\Pi^{ab}_{\mu\nu}(p,m,D),$ obtained by adding the diagrams
in Figs. 3-6. The second step is to compute the large $m$ limit of
the corresponding 3-dimensional result. We shall see that
both limits $D\to 3$ and $m\to \infty$ exist
when taken them in this prescribed order and once we have
summed over all Feynman diagrams. This is in agreement with
refs. [\Blasi] and [\Delduc], where finiteness of
CS theory in the Landau gauge
is proved at any order in pertubation theory.
Let us also point out that in accordance
with the regularization method explained in Sect. 2,
$\Pi^{ab}_{\mu\nu}(p,m,D)$ is the fully regularized
CS vacuum polarization tensor.

At first order in perturbation theory there are three diagrams
that contribute to the polarization tensor (see Fig. 2). Their value
was given in [\Pisarski] for finite $m,$ and the large $m$ limit
was independently computed in\REF\Boss{M. Bos, private
communication.}[\Carmelo,\Semenoff,\Boss]:
$$
\Pi_{\mn}^{(1)ab}(p) = -\,\,{7\over 3}\,\, \xx \,\, \d^{ab}\,\,
      \ee_{\m\n\r}\,p^{\r} \,\, .
\eqn\oneloop
$$
Here we will focuss on the two-loop correction.

We have shown in Sect. 2 that the non-vanishing contributions as
$D\rightarrow 3$ of  the Feynman diagrams in Figs. 3-6 can
be computed by using $D^{ab}_{\mu\nu}(p)$
in eq. \gaugeprop\ as the effective gauge field propagator.
By taking into account that the IR dimension of
$D^{ab}_{\mu\nu}(p)$ is $-1,$ one may readily see that every single
diagram in Figs. 3-6 is IR convergent by power counting at $p\not= 0.$
Thus, no IR singularities occur as $D\rightarrow 3.$ In accordance
with [\Speer], one-loop subintegrations in the diagrams at
hand do not give rise to UV singularities as $D\rightarrow 3$ either,
although they might correspond to superficially UV divergent one-loop
diagrams. However, overall UV singularities do appear
at two loops in individual Feynman diagrams as $D$ approaches 3.
In fact, the two-loop
dimensionally regularized diagrams of Figs. 4, 5 and 6 will develop
UV poles at $x=D-[r/2]$ as $D\rightarrow 3$ if
$x$ is zero or a positive integer, where $r$ is the UV
degree of the integrand and $[r/2]$ is the largest
integer less or equal than $r/2.$ For the diagrams
we are considering $r$ is equal to $6$ so that $x=0.$
Notice that the diagram in Fig. 3 does not exhibit UV divergences, since
it factores into the product of two one-loop diagrams.

Let us now discuss the tensor structure of these overall UV singularities.
The UV degree of the term in $D^{ab}_{\mu\nu}(p)$ involving the
antisymmetric quantity $\ee_{\m\r\n}$ is one unit less than that of
the whole $D^{ab}_{\m\n}(p).$ This very situation is met when studying
the UV degree of the different terms in the three-gauge vertex
$V^{abc}_{\m\r\n}(p,q,r).$ Then, if $\omega_m$ is the overall UV
degree of a diagram at $D=3,$ any integral with $M$ such antisymmetric
objects arising from the diagram will have an
overall UV degree equal to $\omega_m -M.$ The fact that $\omega_m =0$
for the diagrams in Figs. 4, 5 and 6 [see eq. \uvdegree]
implies that the only integrals
exhibiting singularities at $D=3$ are what we might call ``pure"
Yang-Mills integrals, \ie\ those obtained by formally setting to zero
every $\ee_{\m\r\n}$ in the diagram. One may show as well
that any term of these two-loop pure Yang-Mills integrals with at
least one external momentum in the numerator has a negative overall
UV degree. Thus, the only sources of UV singularities at $D=3$
are two-loop integrals having neither $\ee_{\m\r\n}$ nor external
momenta in their numerators. The formal version of these integrals
at $D=n$ is $SO(n)$ invariant
  and we have seen that they have logarithmic superficial degree of
divergence at $D=3.$ Moreover, they are IR finite at $p\not= 0$ and
without UV subdivergences as $D\to 3.$ Hence, their singularity is
a simple pole at $D=3$ independent of $m$ and the external momentum $p$
[\Collins,\Speer]. In addition, they are always multyplied by
$m$ since the mass dimension of the vacuum polarization tensor is one.
We then conclude that the singular contribution coming from the diagrams
in Figs. 4, 5 and 6 has tensor structure
$$
c\,\,{m\over D-3}\,\, g_{\mu\nu} ~,
\eqn\finiteness
$$
where $c$ is a real number whose value depends on the diagram.

We shall show in this section, by explicit computation, that the UV
singularities \finiteness\ cancel when one sums over all
two-loop diagrams contributing to $\Pi^{ab}_{\mu\nu}(p,m,D).$
Let us remind that from the point of view of the na\"\i ve
dimensional regularization of Subsect. 2.1, this was a necessary
step to recover BRS invariance in the 3-dimensional limit.
On the contrary, from the point of view of the BRS-invariant dimensional
regularization defined in Subsect. 2.2, finiteness or cancelation
of singularities is a consequence of BRS invariance at the regularized level.
Indeed, eq. \difference\ and the fact that the singular part of
$\Pi_{\mn}^{(2.1)}$ as $D$ goes to 3 is of the form \finiteness\ imply
that the divergent part of $\Pi_{\mn}^{(2.2)}$ as $D\to 3$ is of the form
$$
m \,\left(\,b_1\,\tilde{g}_{\mn}\, +\, b_2\,\hat{g}_{\mn}\,\right)~,
$$
with
$$
b_1\,=\,{c\over D-3} - {\rm div}\,\{a_1\}~~{\rm and}~~b_2\,=\,
{c\over D-3} - {\rm div}\,\{a_2\}~,
$$
and where ${\rm div}\,\{\cdots\}$ denotes the singular part in the MS sheme
as $D$ approaches
3. Since the dimensionally regularized theory is BRS-invaraint,
$\Pi_{\mn}^{(2.2)}$ is transverse with respect to the external
momentum thus having $b_1=b_2=0.$

We next come to explicit computations.
There are sixteen two-loop diagrams that contribute to the
polarization tensor of the gauge field, see Figs. 3-6. We already know
that poles come from overall logarithmic UV divergences.
A way to check the finiteness of the vacuum polarization
tensor for (3-dimensional) TMYM theory is to collect
all overall divergent integrals from all diagrams, compute them,
and sum the corresponding results. Everything in this process,
but for the evaluation of the integrals, is a question of
algebra and can be performed with the help of an algebraic
language, REDUCE in our case. Here we show with an example how to
compute the integrals and present the final result.

Let us consider the diagrams in Fig. 4.
A typical superficially UV divergent contribution is
$$
m\,g^4\cv^2\,\d^{ab} \meas  {k^2 q^8\,(kq)\,q_{\m}q_{\n}
               \over {D_4(k,q,p,m)}} \,\, , \eqn\div
$$
where the denominator in the integrand is given by
$$
\eqalign{
D_4(k,q,p,m)\, &=\, k^2\,(k^2+m^2)\,(k+q)^2\,[(k+q)^2+m^2] \cr
       &\,\, {\scriptstyle \times}\,\, q^4\,(q^2+m^2)^2
              (q+p)^2\, [(q+p)^2 +m^2] \,\, . \cr }
\eqn\denfish
$$
Note that the integral \div\ has overall UV
logarithmic degree of divergence, in accordance with what has
been said above. To eliminate the dependence on the external
momentum $p$ of the divergences we use the algebraic identities
$$
\eqalign{
{1\over (q+p)^2} =&  {1\over q^2}
          - {2\,pq+p^2\over q^2\,(q+p)^2} \,\, ,\cr
{1\over [(q+p)^2+m^2]}=& {1\over q^2+m^2}
- {2\,pq+p^2\over (q^2+m^2)\,[(q+p)^2+m^2]} \,\, .\cr
}
\eqn\psplit
$$
The integrals arising from the $p\!$-dependent terms on the RHS
of these identities are overall UV convergent at $D=3.$ Moreover,
by inspection we see that they are also IR finite at $p=0.$
The $o\!$-theorem then implies that
they vanish in the limt $m\to\infty,$ since the inequality
\oth\ now takes the form $[n_k+n_q]+n_p > 1,$ with $n_k,\,n_q$ and
$n_p$ the number of $k,q$ and $p$ in the numerator of the corresponding
integrands. Thus, poles are independent of $p$ and to compute them
we can take $p=0$ in $D_4$ so we can rewrite \div\ as
$$
m\,g^4\cv^2\,\d^{ab} \meas { k^2 q^8\,(kq)\,q_{\m}q_{\n}
                    \over {D_4(k,q,0,m)}}
     \,\,\, + \,\,\, {\rm v.t.}  ~~ ,
$$
where ``v.t." stands for finite contributions at $D=3$ which vanish
as $m$ goes to $\infty.$ Rescaling the integration variables
$k\to km,\,\,q\to qm$ and using $SO(D)$ invariance, we have:
$$
{m\over D}\,\,g^4\cv^2\,\d^{ab}\,g_{\mn}
                   \meas {k^2 q^{10}\,(kq) \over {D_4(k,q,0,1)}}
       \,\,\, + \,\,\,{\rm v.t.} ~~  , \eqn\diver
$$
where now the numerator in the integrand is a scalar.
The next step is to use algebraic identities of the type
$$
\eqalign{
&{2\,kq\over (k+q)^2}=1-{q^2\over (k+q)^2}
                         -{k^2 \over (k+q)^2} \,\, ,\cr
&{k^2\over k^2+1}=1-{1\over k^2+1} \,\, ,\cr
&{1\over k^2\,(k^2+1)} ={1\over k^2}-{1\over k^2+1}  \cr
} \eqn\vsplit
$$
to decompose \diver\ into simpler integrals. The
Appendix contains a list of all the integrals found
at the end of this simplification process for the diagrams in Figs. 4-6,
as well as their values. Using the results presented there we get for
the integral in \diver:
$$
{1\over 64 \pi^2} \, \left[ \,- \, {1\over \eps }
      + 1 - \gamma + {325 \over 72}
         + \ln\left({729\pi \over 64}\right)
            + O(\eps) \, \right] ~ ,
\eqn\polefish
$$
where $\gamma$ is Euler-Mascheroni's constant and
$\eps=D-3.$ Notice that \polefish\ exhibits a simple pole at $D=3.$
UV divergent integrals from the diagrams in Figs. 5 and 6 are studied in the
same way. For them we always obtain the same
structure as for \diver, namely a scalar divergent integral times
$m\, g_{\mn}.$ Regarding the diagram of Fig. 3, it factores
into the product of two one-loop diagrams and therefore is
free of poles. After summing over all diagrams we obtain
$$
\Pi_{\mn}^{(2){\rm YM}}(p,m) = - m\, g_\mn
       \left( {1135\over 864} + {9\over 4}\,\ln 2
              - {13\over 16}\, \ln 3 \right)
                      {\left( \xx \right)}^{\!2}\,\,\,
                                + \,\,\, {\rm v.t.} ~~  ,
\eqn\ym
$$
which does not have any singularity.
This proves that the vacuum polarization tensor for TMYM theory is finite
at second order in perturbation theory. The superindex YM in eq. \ym\ is
to remind that this is only the contribution from pure Yang-Mills terms,
or equivalently, terms which by power counting are overall superficially UV
divergent. For simplicity in the notation we have dropped colour indices.

Having shown that the (3-dimensional) topologically massive vacuum
polarization tensor $\Pi_{\mn}^{(2)}(p,m)$ is well-defined,
our next goal is to compute its large $m$ limit. Let us remind that
$\Pi_{\mn}^{(2)}(p,m)$ is obtained  by taking the limit $D\to 3$ of
the sum of the diagrams in Figs. 3-6.
The contribution from pure Yang-Mills integrals is given in eq. \ym. We
still have to calculate the finite contribution $\Pi^{\rm (2)F}_{\mn}(p,m)$
coming from integrals involving one or more
$\ee_{\m\r\n}.$
The na\"\i ve way to evaluate them would be to perform the whole computation
keeping $m$ finite and then take the limit $m\to\infty.$ From the point
of view of the algebra involved, this presents the same degree of
complexity as the computation of the gluon polarization tensor in QCD.
The situation is much worse from the point of view of integration,
since now there are massive denominators. Although this procedure is
still viable at one loop [\Pisarski], it must be given up if we intend
to pursue our computation to higher orders in perturbation theory.
We will use instead the vanishing theorems of Sect. 3.

We begin by applying these theorems to disregard integrals
that go to zero as $m$ approaches infinity. To show how to do this we use
once more the diagrams in Fig. 4 as an example. The most general integral
arising from them is of the form
$$
I(p,m)=\meas {m^{n_m}\, p^{n_p}\, k^{n_k}\, q^{n_q}
          \over {D_4(k,q,p,m)}} \,\, ,  \eqn\exfish
$$
where the denominator is given by \denfish .
The $m\!$-theorem demands the following conditions to
be met for the integral to go to zero as $m\to\infty:$
$$
\eqalign{
{\rm Overall\,\,\,  UV\,\,\, convergence:}\,\,\, & n_k+n_q-14<0 \,\, , \cr
{\rm IR\,\,\, convergence\,\,\, at}\,\,\,p\neq 0:\,\,\,
    & n_k+n_q-2>0 \,\, , \,\, n_k+1 >0\,\,
                       , \,\, n_q-1>0\,\, , \cr
{\rm Condition\,\,\,\mth\ :}\,\,\,
    & n_k+n_q+n_m-14<0\,\, , \,\, n_m-10<0\,\, , \cr
    & n_k+n_m-11<0 \,\, , \,\, n_q+n_m-15<0 \,\, , \cr}
$$
where we have have already used that UV convergence along one-loop
subintegrals can be dropped, as discussed at the end of Sect. 3.
The Feynman rules as applied to the diagrams we are considering imply
that the only non-trivially satisfied conditions are
$$
n_k+n_q+n_m-14<0  ~~{\rm and}~~ \quad n_k+n_m-11 <0 \,\, .
\eqn\mthfish
$$
Analogously, one can obtain that the $o\!$-theorems demands:
$$
n_k+n_q-14<0 \,\, , \,\, n_k+n_q-4>0 \,\, , \,\,
n_q-3>0 \,\, , \,\,
[n_k+n_q]+14-n_k-n_q-n_m>0 \,\, . \eqn\othfish
$$
We now use the cuts \mthfish\ and \othfish\ to keep only those integrals
\exfish\ which do not vanish in the limit $m\to\infty.$ In this way we are
left with a number of integrals that have to be computed. In what follows
we give two examples of such integrals and show how to evaluate them. Again
the $m\!$- and $o\!$-theorems will play a very important part.

Let us first consider the integral
$$
I_\mn(p)= \meas {m^7 k^4 q^2 q_\mu q_\nu \over D_4(k,q,p,m)} ~ ,
\eqn\exboth
$$
with the denominator in the integrand given by \denfish. Note that
$n_k+n_q+n_m=15$ and $[n_k+n_q]+14-n_k-n_q-n_m=0$ so that the theorems did
not apply. This integral is IR finite at $p=0$ for $D=3$ so we must expect
a contribution from $p=0,$ which will be of order $m,$ plus subleading
contributions of order $m^0.$ Using the identities \psplit\ and the
$o\!$-theorem it is not difficult to see that
$$
I_{\mn}(p) = I_{\mn}(0) + I'_{\mn}(p)
          +  {\rm v.t.} ~~  ,
$$
where
$$
I_\mn'(p)= - \meas {m^7 k^4 q^2 \,(2\,pq + p^2)\, q_\m q_\n
               \over D'_4(k,q,p,m)} \,\,
$$
and the denominator $D'_4$ reads
$$
D'_4(k,q,p,m) = k^2 \,(k^2+m^2) \, (k+q)^2 \, [(k+q)^2+m^2]
                (p+q)^2\, q^6 \,(q^2+m^2)^3 \,\, .
$$
The integral $I_{\mn}(0)$ is of the same type as those that appeared
when studying the pure Yang-Mills sector, with the difference that
this one is finite at $D=3.$ It can be computed in the same way,
the result being
$$
I_{\mn}(0) = m g_{\mn}\,\,{1\over 48\,\pi^2}\,
  \left( {143\over 288} - 2\,\ln 2 + \ln 3 \right) ~ .
$$
Regarding $I'_{\mn}(p),$ it can be further
simplified. To decouple integration over $k$ from that over $q$ we use
the algebraic identities
$$
\eqalign{
{1\over (k+q)^2}=& {1\over k^2}
          -{2\,kq+q^2 \over k^2\,(k+q)^2} \,\, ,\cr
{1\over (k+q)^2+m^2}=& {1\over k^2+m^2}
-{2\,kq+q^2 \over (k^2+m^2)\,[(k+q)^2+m^2]} \,\, . \cr }
\eqn\qsplit
$$
The terms with dependence on $q$ on the RHS of \qsplit\ give zero
contribution to $I'_{\mn}(p)$ when $m\to\infty$ by application of the
$o\!$-theorem. We thus get for $I'_{\mn}(p)$ the product of two one-loop
integrals:
$$
I'_{\mn}(p) = - \meask {m\over (k^2+m^2)^2}
            \measq {m^6\!q^2\,(2\,pq+p^2)\,q_{\m}q_{\n}
                  \over (q+p)^2\,q^6\,(q^2+m^2)^3}\,\,\,
          +\,\,\, {\rm v.t.} ~~  .
$$
Rescaling $k \to km$ we concentrate all the dependence on $m$ on the
integral over $q.$ Using then identities of the type \vsplit\ and the
$o\!$-theorem we finally get that
$$
\eqalign{
I'_{\mn}(p) & = \meask {1\over (k^2+1)^2}
                  \measq {q_{\m}q_{\n} \over q^2\,(q+p)^2}
                      \,\,\, +\,\,\, {\rm v.t.}   \cropen{12pt}
     & = -\, {1\over 512\pi\sqrt{p^2}} \,
            \left( p^2\, g_{\mn} - 3\,p_\m p_\n \right)
                 \,\,\, + \,\,\, {\rm v.t.} ~~ , \cr}
$$
which is of order $m^0.$ Another typical example of an integral
with non-zero limit $m\to\infty$ is
$$
\meas
{m^9 k^2 q^2\,p_{\m}q_{\n} \over D_4(k,q,p,m)}
          =  -\, {p_\m p_\n \over 128 \pi \sqrt{p^2}}
\,\,\, + \,\,\, {\rm v.t.} ~~ .
$$
In this case the integral had $n_k+n_q+n_m=14$ and was IR divergent at
$p=0$ for $D=3$ so the theorems did not apply. Nevertheless, its large
$m$ limit has a simple expression.  Note that both examples yield
non-local terms.

All the integrals from the diagrams in Figs. 3-6 that do not vanish when
$m$ goes to $\infty$ can be computed in the same way as these two examples.
When we sum over diagrams, non-localities cancel and the contribution
to the polarization tensor from terms with one or more $\ee_{\m\n\r}$
takes the form:
$$
\eqalign{
\Pi_{\mn}^{\rm (2) F} & = m\, g_\mn\,\,
       \left( {1135\over 864} + {9\over 4}\,\ln 2
              - {13\over 16}\, \ln 3 \right)
                   {\left( \xx \right)}^{\!2} \cr
     &+ \ee_{\m\r\n}\,p^{\r}\,\,
       \left( {265\over 36} + {44\over 3}\,\ln 2
              - {63\over 4}\, \ln 3 \right)
                   {\left( \xx \right)}^{\!2}
           \,\,\, + \,\,\, {\rm v.t.} ~~  . \cr }
\eqn\pidoscs
$$
Combining this result with the contribution \ym, we finally get
$$
\Pi_{\mn}^{(2)ab}(p,m) = \ee_{\m\r\n}\,p^{\r}\,\,\d^{ab} ~
      {265+L\over 36} ~ {\left( \xx \right)}^{\!2} ~ ,
\eqn\twoloop
$$
where we have reinserted colour indices and introduced the constant
$$
L=528 \,\ln 2 - 567 \,\ln 3 \,\, . \eqn\constant
$$
We thus see that when the limit $m\to\infty$ is taken no infinities
appear. Eq. \twoloop\ gives the $g^4\!$-correction to the CS gauge field
polarization tensor and will be used in Sect. 8 to construct the effective
action.

Notice that since limit $m\to \infty$ prserves BRS
invariance, the regularized CS vacuum polarization tensor must
be transverse. Hence, if the $mg_{\mu\nu}$ terms in
$\Pi^{\rm (2) YM}_{\mn}$ in eq. \ym\ and $\Pi^{\rm (2) F}_{\mu\nu}$ in eq.
\pidoscs\ did not cancel one another, transversality would be lost.
Finally, let us point out that the absence of non-local large $m$ divergent
terms is due to the lack of UV subdivergences after summing over
all two-loop diagrams, or in other words, to UV finiteness of one-loop
1PI functions as $m\to\infty.$

{\bf\chapter{The ghost self-energy and the ghost-ghost-external
field vertex}}

One of our aims is to construct the local part of the two-loop effective
action for CS theory. We will do this in the next section. To determine
the coefficients of the terms entering in the effective action
we will need to know three Green functions.
We have already taken one of them to be the gauge field propagator,
which was studied in Sect. 5. For the other
two we chose the ghost propagator and the ghost-ghost-external
field vertex, whose results we give here.

Let us first consider the ghost propagator. We want to
compute the ghost self-energy $\Omega^{ab}(p)$ up to second order
in perturbation theory.
At the first perturbative order there is only one
diagram that contributes to the self-energy, see Fig. 7.
Its contribution in the limit $m\to\infty$
can be easily computed [\Pisarski,\Semenoff ],
the result being
$$
\Omega^{(1)\,ab}\,(p) = -{2\over 3}\,\, \xx
            \, p^2\,\delta^{ab} \,\, .
\eqn\ghostone
$$
At second order there are six diagrams that contribute, see Fig. 8.
We use the theorems of Sect. 3 to keep
only those integrals that do not vanish as $m$ goes to $\infty.$
Their evaluation in the limit $m\to\infty$ goes along the lines explained
in the last section. As a matter of fact, it happens that all the integrals
one has to evaluate are among those studied in Sect. 5. Here we limit
ourselves to give the result:
$$
\Omega^{(2)\,ab}\,(p)= - \,{{169+L}\over 72}\,\,
         {\left(\xx\right)}^{\!2}\, p^2\,\delta^{ab} \,\, ,
\eqn\ghosttwo
$$
where $L$ is as in eq. \constant.

It is worth noticing that despite power counting as applied to CS
theory predicts linear UV divergences, the results in \ghostone\
and \ghosttwo\ are finite, in agreement with [\Blasi,\Delduc].
Moreover, first and second order corrections to the ghost self-energy
in TMYM theory are also finite: they are given in terms of finite integrals
whose large $m$ limit precisely gives \ghostone\ and \ghosttwo. Again, this
is in contrast with the logarithmic UV divergences predicted
by power counting for TMYM one-loop diagrams.

We next study the ghost-ghost-external field vertex $H^a c^b c^c.$
At one loop there is only one Feynman diagram
(see Fig. 9). It is not difficult to shwow that its contribution to the
vertex vanishes in the $m\to\infty$ limit,
$$
V^{(1)\,\,abc}\,(p_1,p_2)=0 \,\,\, + \,\,\, {\rm v.t.}~~~ .
\eqn\extone
$$
As for second order corrections, Fig. 10 depicts all two-loop
diagrams that contribute. The way to compute them is analogous to that of
calculating the ghost self-energy and gauge field vacuum polarization
tensor. The only difference is that now it is enough to use the
$m\!$-theorem. Let us illustrate the method for the diagram in Fig 10(a).
We get for its contribution:
$$
{1\over 4}\,\,g^5\,\cv^2\,f^{abc}
     \int \, {d^3\!k\over (2\pi)^3}\,\,
             {d^3\!q\over (2\pi)^3}\,\,\,{{N(k,q,p_1,p_2,m)}
                     \over D_{10}(k,q,p_1,p_2,m)} \,\, ,
\eqn\verex
$$
where the numerator and denominator of the integrand are given by
$$
\eqalign{ N(k,q,p_1,p_2,m)\, & =
       m^2\,p_1^{\m}\,p_2^{\n}\,q^{\r}\,(q+p_1+p_2)^{\s} \cr
    &\,\, {\scriptstyle \times} \, \left[ m\,\ee_{\m\t\n}\,(q+p_1)^{\t} +
       (q+p_1)^2\,\d_{\m\n} - (q+p_1)_{\m}\,(q+p_1)_{\n} \right] \cr
    &\,\, {\scriptstyle \times} \, \left[ m\,\ee_{\r\l\s}\,k^{\l} +
       k^2\,\d_{\r\s} - k_{\r}k_{\s} \right] ~ ,\cr}
$$
and
$$
\eqalign{
D_{10}(k,q,p_1,p_2,m)\, &= \,q^2\,(q+p_1)^2\,
     \left[ (q+p_1)^2+m^2 \right] \,(q+p_1+p_2)^2\cr
    &\,\, {\scriptstyle \times} \,k^2\,(k^2+m^2)\,(k+q)^2\,
         (k+q+p_1+p_2)^2  \,\, .\cr} $$
For non-zero external momenta, the diagram is by power counting finite.
The $m\!$-theorem then demands that for \verex\ to vanish as $m\to\infty,$
condition \mth\ must hold. But condition \mth\ requires the following
inequalities to be satisfied simultaneously:
$$
n_k+n_q+n_m-10<0 \,\, , \,\, n_k+n_m-7<0 \,\, , \,\,
n_q+m_m-11<0 \,\, , \,\, n_m-4<0\,\, ,
$$
where as usual $n_k,n_q$ and $n_m$ denote the number of
$k{\rm 's},\>q{\rm 's}$ and $m{\rm 's}$ in the numerator of the
integrand. The only term that violates these requirements is
$$
m^4\, p_1^{\m}\,p_2^{\n}\,q^{\r}\,(p_1+p_2)^{\s}\,
  \ee_{\m\t\n}\,q^{\t}\,\ee_{\r\l\s}\,k^{\l} \,\, ,$$
for which $n_m=4.$ Now, integration over $k$ yields a linear combination
of $q^{\l}$ and $(p_1+p_2)^{\l}$ that together with the
$\ee_{\r\l\s}q^{\r}\,(p_1+p_2)^{\s}$ gives zero. Proceeding in a similar
way we obtain for the other diagrams the following. The colour algebra of
the diagram in Fig. 10.(b) is already zero. The $m\!$-theorem implies that
each one of the diagrams in Figs. 10(c)-10(e) on its own vanishes as $m$
goes to $\infty.$ Finally, the diagrams in Figs. 10(f) and 10(g) on the
one hand, and Figs. 10(h) and 10(i) on the other, combine to give
zero contribution for $m\to\infty.$ We thus obtain that
$$
V^{(2)\,\,abc}\,(p_1,p_2)=0\,\, .\eqn\exttwo
$$
This result and \ghosttwo\ will be used in the Sect. 8.

{\bf\chapter{Perturbative finiteness of TMYM theory}}

Here we collect some results from previous sections that
put together imply perturbative finiteness for TMYM theory.
It follows from eq. \uvdegree\ that the only 1PI Green functions
that are not finite by power counting are:

{\leftskip=1 true cm \rightskip=1 true cm
\noindent a) The gauge field one-point function $<\!A^a_\m\!>\,,$ with
eq. \uvdegree\ predicting quadratic, linear and logarithmic UV
divergences at one, two and three loops respectively.

\noindent b) The vacuum polarization tensor, for which power counting
gives linear UV divergences at one loop and logarithmic at two.

\noindent c) The ghost self-energy and the vertex
$<\!A^a_\m A^b_\n A^c_\r\!>\,,$ power counting yielding one-loop
logarithmic UV divergences.
\par}

\noindent To give a well-defined meaning to these Green functions we have
introduced dimensional regularization in Sect. 2.

As shown in Sect. 4, any 1PI diagram contributing to
$<\!A^a_\m\!>$ is zero by colour algebra upon regularization. This leaves
us with only b) and c) above as sources of UV divergences. In Sect. 2 we
have seen that the limit $D\to 3$ of every dimensionally
regularized one-loop 1PI Feynman diagram is free of singularties, despite
power counting might predict divergences. This 
ensures finiteness at one
loop. Furthermore, being the limit $D\to 3$ finite at one loop, it can be
regarded as defining a renormalization scheme at this order. Thus,
two-loop UV divergences may only come from the vacuum polarization tensor.
In Sect. 5 we have seen that although individual
two-loop 1PI diagrams contributing to the polarization tensor have poles
at $D=3,$ the latter cancel when summing over diagrams thus providing a
finite polarization tensor. Hence,
no UV divergences arise at second order in perturbation theory. Again,
the limit $D\to 3$ defines a renormalization scheme at this
order. This, combined with finiteness by power counting at higher loops
implies that TMYM theory is finite at any perturbative order.

Finiteness at one loop was proved in [\Deser,\Pisarski]. At higher loops,
though expected, has remained unproved. Also in [\Deser,\Pisarski] explicit
expressions for one-loop radiative corrections can be found. Finding
compact expressions for such corrections beyond one loop seems today
a task beyond human (and computer) capability. In this paper we have
computed the large $m$ limit of some 1PI functions at two loops

{\bf\chapter{The bare effective action}}

In this section we compute the first and second order radiative corrections
to the bare CS effective action $\gm$ up to order three in the fields.
We shall work in the Landau gauge, as we have done so far. The bare effective
action is defined to be
$$
\gm\, =\, \gm (A,c,{\bar c},b, J, H; k)\,=\,
      \lim_{m\to \infty}\,\,\lim_{D\to 3}\,\,
             \gm (A,c,{\bar c},b,J,H; k; m,D) ~ ,
\eqn\effaction
$$
wherever the previous sequence of limits exists. We will assume for the time
being that the above double limit is indeed finite up to second order in
perturbation theory and postpone its proof until the end of this section.
This amounts to assuming UV finiteness of $\gm$ up to two loops. Note
that this is in agreement with the general finiteness proofs
in [\Blasi] and [\Delduc]. In \effaction, the integrated functional
$\gm (A,c,{\bar c},b,J,H;k;m,D)$ is the generating functional of the amputated
1PI functions regularized through any of the regularization methods in Sect.
2. Let us recall that both methods yield the same result in the limit
$D\to 3.$

Throughout this section we will write $\gm$ as a function of $k\!>\!0,$
the classical or bare CS parameter, rather than as a function of the
classical or bare coupling constant $g=(4\pi/k)^{1/2}.$ This can be
accomplished by introducing the following scalings of the fields:
$$
A^{\m a}\to {1\over g} \,A^{\m a}~~,~~
b^a\to g b^a ~~,~~
c^a\to {1\over g} \,c^a ~~,~~
\bar{c}^a\to g \bar{c}^a ~~,~~
J^{\m a}\to g J^{\m a} ~~,~~
H^a\to g H^a  \,\, .
$$

In computing $\gm$ we will use the BRS identities, the values of the
1PI functions calculated in previous sections and the
following result. Both the
one-loop and the two-loop terms in the effective action are local integrated
functionals of the fields, provided they correspond to 1PI functions with
less than than four fields. This statement is a consequence of two facts:
({\it i}) the regularization method in [\Guadagnini] yields vanishing
radiative corrections up to two loops for the 1PI functions here considered,
and ({\it ii}) two renormalization prescriptions for the effective action
that agree at $n$ loops can only differ by a local functional
at $n+1$ loops\REF\Hepp{K. Hepp, {\it Th\'eorie de la renormalisation},
Lectures Notes in Physics vol. 2, Springer-Verlag, Berlin (1969).}[\Hepp].

As for the BRS identities, we are going to write them in terms
of the functional $\gmb:$
$$
\gmb\,=\,\gm + \idx\,\, b^a\partial A^a \,\, ,
\eqn\baraction
$$
with $\gm$ as in eq. \effaction. The functional $\gmb$ satisfies the
following equations:
$$
{{\d{\gmb}} \over {\d b^a}} \, = \,0 ~,~~~
\left( \partial_{\m}\,\, \dj - \dcb \right)
      {\gmb} \,=\, 0 \,\, , \eqn\constrains
$$
$$
\idx \left( \dgmba\,\dgmbj + \dgmbc \, \dgmbH \right)
        \,=\,0 \,\, . \eqn\brseq
$$
Let us discuss why  eqs. \constrains\ and \brseq\ hold.
The two equations in
\constrains\ are the Landau gauge-fixing condition and the antighost
equation of motion respectively. They hold for the regularized
$\gmb (A,c,{\bar c},b,J,H; k;m, D)$ obtained through any of the
regularization methods in Sect. 2 and the $D\!$-dimensional analogue of
eq. \baraction. Eq. \brseq\ is the BRS equation. This equation is satisfied
by the regularized effective action $\gmb (m,D)$ that the second
regularization method in Sect. 2 provides (see Subsect. 2.2). Let us
recall that this method is explicitly BRS-invariant and
that eq. \brseq\ is the equation ruling
BRS invariance for the quantum theory. On the other hand, the
first regularization method in Sect. 2 (see Subsect. 2.1)
yields a regularized effective action $\gmb(m,D)$ that does
not satisfy the BRS equation. However, we have shown that the terms
breaking the BRS symmetry go to zero as $D\rightarrow 3.$
Moreover, both regularization methods yield the same effective
action in the limit $D\to 3.$ We then conclude that $\gmb$ in eq. \baraction\
does satisfy the set of equations above, hence that it is BRS-invariant.

The first equation \constrains\ implies that $\gmb$ does not depend on
$b^a.$ In turn, the antighost equation leads to the conclusion
that $J^a_{\mu}$ and ${\bar c}^a$ always occur through the combination
$$
G^a_{\m}(x)\,=\,J^a_{\m}(x) -\partial_{\m} {\bar c}^a (x)\,\, .
\eqn\combination
$$
Thus, the functional $\gmb$ should be understood as a function of
$A^a_{\m},\,\, G^a_{\m},\,\,c^a$ and $H^a.$ As an integrated functional,
$\gmb$ has mass dimension three and ghost number
zero. The fields $A^a_{\m},\,\,G^a_{\m},\,\,c^a$ and $H^a$ all have mass
dimesion $1$ and ghost number $1,\,\,-1,\,\,1$ and $-2$
respectively. From these mass dimensions and ghost numbers it follows that
contributions to $\gmb$ quadratic or cubic in the fields are local,
whereas contributions quartic or higher are purely non-local
functionals, \ie\ with no local part. Recall CS theory in the
Landau gauge does not involve any dimensionful parameter.

As for eq. \brseq, it establishes relations among the coefficients of
the 1PI functions contributing to $\gmb.$ These relations, toghether
with the values of the 1PI functions we have computed in previous sections,
fix completely the local part of $\gmb(A,c,G,H;k)$ up to two loops.
To see this we introduce a loop-wise expansion for $\gmb:$
$$
\gmb \,=\, \sum_{n=0}^{\infty} \hbar^n \gmbn ~ ,
\eqn\loopwise
$$
where $\gmbz$ is the tree-level effective action minus the gauge-fixing
term, or more precisely:
$$
\gmbz\,=\,  S_{CS} + \idx
\left( G^{a\m}D^{ab}_{\m}c^b-{1\over 2}f^{abc}H^a c^b c^c \right) ~ .
\eqn\treeaction
$$  
The symbol $S_{CS}$ stands for the CS classical action in terms of $k,$
$$
S_{CS}\,=\,-\,{ik\over 4\pi}\idx\,\ee^{\m\r\n}
    \left( \,{1\over 2}\,A^a_{\m}\partial_{\r}A^a_{\n}
        + {1\over 3!}\,f^{abc}\,A^a_{\m}A^b_{\r}A^c_{\n}\,\right) ~ .
$$
Substitution of eq. \loopwise\ in eq. \brseq\ gives the following
equations for the one-loop $\gmbo$ and two-loop $\gmbt$ contributions
to $\gmb:$
$$
\dd\,\gmbo \, =\,0 \,\, ,
\eqn\onebrs
$$
and
$$
\idx \left[ \dgmboa\,\dgmboG + \dgmboc\,\dgmboH \right]
          + \dd\,\gmbt \, =\,0 \,\, .
\eqn\twobrs
$$
Here $\dd$ is the Slavnov-Taylor operator,
$$
\dd \,=\, \idx \left[ \dgmbza\, \dG + \dgmbzG \, \da
               + \dgmbzc \, \dH + \dgmbzH \, \dc \right]\,\, .
\eqn\SToper
$$
As is well-known, $\dd$ is nilpotent, $\dd^2\,=\,0.$ Notice that in passing
from eq. \brseq\ to eqs. \onebrs\ and \twobrs\ we have taken into account
that
$$
\dgmbj\,=\,\dgmbG\,\, .
$$

Eq. \onebrs\ imposes some constraints upon the local
part $\gmbo_{\rm local}$ of $\gmbo.$ The most general way to find them
would be to solve the equation
$$
\dd \,W\,=\,0 ~
\eqn\mastereq
$$
over the space of integrated functionals $W$ of mass dimension three and
ghost number zero that depend on $A^a_{\m},\,\,c^a,\,\,G^a_{\m}$
and $H^a.$ In principle, $W$ has local and non-local contributions
and, in perturbation theory, can be expressed as the sum
$$
W(A,G,c,H)\,=\,\sum_{i=2}^{\infty}W_i\, (A,G,c,H) ~,
\eqn\filter
$$
with the index $i$ counting the number of fields in $W_i.$ Note that
we have restricted the sum to $i\geq 2,$ since contributions to the
effective action $\gmb$ at any order in perturbation theory, and in
particular to $\gmbo,$ are at least quadratic in the fields, as shown in
Sect. 4. Furthermore, as we have already mentioned, contributions of order
two and three in the fields are local, and contributions of order four or
higher are non-local. This means that $W_2$ and $W_3$ are local functionals
and that $W_i\,\,(i\geq 4)$ is purely non-local. Regarding the Slavnov-Taylor
operator $\dd,$ it is convenient to split it also into two terms, each one
of them of a given order in the number of fields. To do this we decompose
$\gmbz$ in the sum of its quadratic $\gmbz_2$ and
cubic $\gmbz_3$ parts in the fields,
$$
\gmbz (A,c,G,H)\,=\,\gmbz_2 (A,c,G)\,+\,\gmbz_3 (A,c,G,H) ~.
\eqn\splitgmo
$$
Calling now $\dd_0$ and $\dd_1$ to the Slavnov-Taylor
operators for $\gmbz_2$ and $\gmbz_3$ respectively, we have that
$$
\dd\,=\,\dd_0\,+\,\dd_1~,
\eqn\zeroplusone
$$
with
$$
\dd_0^2\,=\,\dd_1^2\,=\,\{\dd_0 ,\dd_1\}\,=\,0 ~.
$$

We are interested in the most general $W_2$ and $W_3$ entering
in the solution $W$ of eq. \mastereq, since our $\gmbo_{\rm local}$
is the sum of two specific $W_2$
and $W_3.$ Eq. \mastereq\ leads to an infinite number of coupled equations
for the functionals $W_i.$ However, only three of them involve $W_2$ and
$W_3,$ namely:
$$
\dd_0\,W_2\,=\,0~~,~~\dd_0\,W_3\,+\,\dd_1\,W_2\,=\,0
\eqn\nice
$$
and
$$
\dd_1\,W_3\,+\,\dd_0\,W_4\,=\,0 ~ .
\eqn\nasty
$$
The last equation containing the purely non-local contribution $W_4.$
It can be readily seen after some algebra that the most general integrated
functional $W_{\rm local}\,=\,W_2+ W_3,$ with $W_2$ and $W_3$ solutions of
eq. \nice, is
$$
\eqalign{
W_{\rm local} \,=\, & -\>{ik \over 4\pi} \idx \,\,\ee^{\m\r\n}
       \left[ {1\over 2}\,\left( w_1+2\,w_2 \right)\,
             A_{\m}^a \partial_{\r} A_{\n}^a +
       {1\over {3!}}\, \left( w_1+3\,w_2 \right)\,f^{abc}\,
             A_{\m}^a\,A_{\r}^b\,A_{\n}^c \right] \cr
       & + \idx\, \left[\, - w_2\, G^a_{\m}\partial^{\m}c^a
             + w_3\, G^a_{\m}{\left( D^{\m}c \right)}^a
             - {w_3\over 2}\,f^{abc}\,H^ac^bc^c \right] ~ , \cr}
\eqn\workedsolution
$$
where $w_1,\,\,w_2$ and $w_3$ are arbitrary coefficients. Actually, it is
easy to check that the first equation in \nice\ holds for any functional
$W_2$ of mass dimension three and ghost number zero. The functional
$W_{\rm local}$ can be recast into the form
$$
W_{\rm local}\, (A,G,c,H)\,=\,w_1\, S_{CS}\,+\,\dd X ~ ,
\eqn\solution
$$
with
$$
X\,=\, \idx\, \left( w_2\,G^a_{\m} A^{a\m}\,-\,w_3\,H^a c^a\right)\,\, .
\eqn\generalX
$$
Now, it is well-known [\Blasi] that \solution\ is the most general
solution of the equation
$$
\dd\,W_{\rm local}\,=\,0 ~
\eqn\blasito
$$
over the space of integrated local functionals of the fields
$A^a_{\m},\,\,G^a_{\m},\,\,c^a$ and $H^a$ with mass dimension three and
ghost number zero. From eqs. \blasito\ and  \zeroplusone\ it then follows
that $\dd_1 W_3$ vanishes for any $W_3$ solving eq. \nice.
This, along with eq. \nasty,
implies that $\dd_0 W_4 = 0$ so that the purely non-local sector of $W,$
defined in eq. \filter, decouples from the local sector as far as BRS
invariance is concerned. This is also a property of $\gmbo_{\rm local},$
for it belongs to the space of functionals spanned by $W_{\rm local}.$

To find the values of $w_1,\,\,w_2$ and $w_3$ that enter in
$\gmbo_{\rm local}$ we exploit that the polarization tensor
$\Pi_{\mn}^{ab}(p),$ the ghost self-energy $\Omega^{ab}(p)$ and
the external vertex $V^{abc}(p_1,p_2)$ are generated by functional
derivatives of $\gmb$ with respect to the fields:
$$
\idp \, \Pi_{\mn}^{ab}(p) \,  = \,
     -\,\, {{\d^2\gmb} \over {\d A^{\m a}(x) \d A^{\n b}(y)}}
           {\Biggl\arrowvert} _{A=0} \,\, ,
$$
$$
\idp \, \Omega^{ab}(p) \, = \,
     -\,\, {{\d^2\gmb} \over {\d c^a(x) \d \bar{c}^b(y)}}
           {\Biggl\arrowvert} _{\bar{c}=c=0} \,\, ,
$$
$$
\idpp \, V^{abc}(p_1,p_2) \, = \,
     -\,\, {{\d^3\gmb} \over {\d H^a(z) \d c^b(x) \d \bar{c}^c(y)}}
           {\Biggl\arrowvert} _{H=\bar{c}=c=0} \,\, .
$$
By replacing $\gmb$ with $W_{\rm local}$ and taking into consideration
the values of $\Pi_{\mn}^{ab}(p),\,\,\Omega^{ab}(p)$ and $V^{abc}(p_1,p_2)$
in eqs. \oneloop, \ghostone\ and \extone, we obtain:
$$
w_1\,=\,w_1^{(1)}\,=\,\yy~~,~~w_2\,=\,
w_2^{(1)}\,=\,{2\over 3}\yy~~,~~w_3\,=\,w_3^{(1)}=0 ~.
\eqn\wone
$$
Note that these values, together with $W_{\rm local},$ fix completely the
local contribution $\gmbo_{\rm local}$ to $\gmbo.$ The superscripts
in $w_1^{(1)},\,\, w_2^{(1)}$ and $w_3^{(1)}$ stand for ``one-loop''.

We next compute the local part $\gmbt_{\rm local}$
of the two-loop contribution $\gmbt$ to the bare effective action.
Again, terms with more than three fields are purely
non-local, whereas those involving less than four fields
are local. Thus, if $\gmbt_2$ and $\gmbt_3$ denote respectively the
two- and three-field contributions to $\gmbt,$ we have
$$
\gmbt_{\rm local}\,=\,\gmbt_2\,+\,\gmbt_3\,\,.
\eqn\twosplit
$$
In what follows we find $\gmbt_{\rm local}$ by solving eq. \twobrs\ up to
order three in the number of fields. To do so we introduce the functional
$Y_3:$
$$
Y_3\,=\,\idx \left[ \dgmbola\,\dgmbolG + \dgmbolc\,\dgmbolH \right] ~ ,
\eqn\ythree
$$
with $\gmbo_{\rm local}$ given by eq. \workedsolution\ with coefficients
\wone. It is a matter of some algebra to see that the functional $Y_3$
takes the form:
$$
Y_3\,=\,{ik\over 4\pi}\,\,{\left( \yy\right)}^{\! 2} \,\,
             \idx \,\, \ee^{\m\r\n} \, f^{abc}\,
                A_{\m}^a A_{\r}^b \partial_{\n}c^c \,\, .
$$
The key point now is to observe that $Y_3$ can be written as
$$
Y_3\, =\, \dd_1 \Upsilon_2 ~ ,
$$
where the functional $\Upsilon_2$ reads
$$
\Upsilon_2\,=\, - \>{ik\over 4\pi} \,\, {\left( \yy\right)}^{\! 2} \idx \,\,
         \ee^{\m\r\n} \, A_\m^a\partial_{\r} A_{\n}^a \,\, .
$$
Eq. \twobrs\ then implies that $\gmbt_2$ and $\gmbt_3$ satisfy
the following two equations:
$$
\dd_0\,\gmbt_2\,=\,0~~~,~~~
\dd_0\,\gmbt_3\,+\,\dd_1\,\left[\Upsilon_2\,+\,\gmbt_2\right]\,=\,0 ~ ,
\eqn\nicetwo
$$
where we have used that $\dd_0 \Upsilon_2$ vanishes.
Eqs. \nicetwo\ are again of the type \nice, with
$W_2\,=\,\Upsilon_2\,+\,\gmbt_2$ and $W_3\,=\,\gmbt_3,$ and
are to be solved over the same space of functionals.
We thus conclude that $\Upsilon_2\,+\,\gmbt_{\rm local}$
is of the form \workedsolution, or equivalently \solution. The
coefficients $w_1,\,\,w_2$ and $w_3$ can be calculated
in the same way as before; for them we obtain:
$$
w_1\,=\,w_1^{(2)}\,=\,{14\over 3}\, {\left(\yy\right)}^{\!2} \quad , \quad
w_2\,=\,w_2^{(2)}\,=\,{{169+L}\over 72}\, {\left( \yy\right)}^{\!2} \quad,\quad
w_3\,=\,w_3^{(2)}\,=\,0\,\, ,
$$
with $L$ as in eq. \constant. Putting everything together we have that the
local part of the effective action $\gm_{\rm local}$ up to second order in
perturbation theory is:
$$
\eqalign{
\gm_{\rm local} \,=\, & -\,{ik\over 4\pi} \idx\,\, \ee^{\m\r\n} \left(
       {1\over 2} \, R_1\,A_{\m}^a \partial_{\r} A_{\n}^a +
       {1\over{3!}}\, R_2\,f^{abc} A_{\m}^a\,A_{\r}^b\,A_{\n}^c
                                                      \right) \cr
& + \idx \left[\,G_\m^a \left( R_3\,\partial_{\m}c^a
                                  - f^{abc}c^bA^{\m c} \right)
       - b^a \partial A^a - {1\over 2}\,f^{abc}\, H^ac^bc^c \right] ~ . \cr}
\eqn\effectotal
$$
where $R_1, R_2$ and $R_3$ are functions of $k^{-1}$ given by
$$
\eqalign{
R_1\,&=\, 1 + {7\over 3}\,\, \yy
      + {{265+L}\over 36} \,{\left( \yy\right)}^{\!2}~ ,\crr
R_2\,&=\, 1 + 3 \,\,\yy
      + {{281+L}\over 24} \,{\left(\yy\right)}^{\!2} ~ ,\crr
R_3\,&=\, 1 - {2\over 3} \,\,\yy
      - {{169+L}\over 72} \,{\left(\yy\right)}^{\!2} \,\, .\cr}
$$

To close this section we show that the double limit in eq. \effaction\
exists up to two loops. We have seen in Sect. 7 that the
limit $D\to 3$ of the dimensionally regularized TMYM effective action
$$
\gm(A,b,G,c,H;k;m,D)\,=\,
\gmb (A,G,c,H; k; m,D)\,-\,\int\, d^D\!x\, b^a\partial A^a\,\, ,
$$
is finite at any order in perturbation theory. Hence, the limit
$D\to 3$ in \effaction\ exists. Moreover, we know that both dimensional
regularization methods in Sect. 2 yield the same result as $D$ goes
to 3 and that this limit is BRS-invariant. This means that the bare
TMYM action $\gmb (m)$ in \baraction\ satisfies
the BRS equation \brseq. In particular, its one-loop contribution must
satisfy eq. \mastereq, and so must the corresponding one-loop divergent
part as $m$ goes to infinty (if any). Thus, if there were a one-loop
large $m$ divergent contribution, it would be of the form \workedsolution,
with $w_1,\,\, w_2$ and $w_3$ carrying large $m$ singularities.
However, these coefficients can not be but zero, for the vacuum
polarization tensor, the ghost self-energy and the $Hcc$ vertex are
finite as $m$ goes to infinity. We then infer that the double limit in eq.
\effaction\ exists at one loop.
This, together with eq. \twobrs, implies that the would-be two-loop large $m$
divergent contribution to $\gmb (m)$ would satisfy eq. \mastereq\ again.
The same argument as for the one-loop order leads
to large $m$ finiteness at two loops.
Thus, our regularization method provides an UV finite CS effective
action up to second order in perturbation theory in the least.

{\bf\chapter{Absence of two-loop corrections to Witten's shift.}}

The purpose of this section is to show that two-loop contributions to
the bare effective action in eq. \effectotal\ do not modify Witten's shift
$k\to k+\cv,$ where $k>0$ is the classical CS parameter
[\Witten]. We will see in particular that two-loop contributions
correspond to a cohomologically trivial term in the sense of the
cohomology of the Slavnov-Taylor operator. Hence, they are equivalent to
a wave function renormalization of the fields and can be set
to zero by an appropriate rescaling of the latter. Let us recall that in
quantum field theory [\Collins] the values of the
observables are not modified by finite rescalings of the fields.
Furthermore, the need of introducing wave function renormalizations to
bring out the physical effects of radiative corrections
to 1PI functions is a common feature in quantum field theory.
Think, for instance, of the computation of the beta function from vertex
radiative corrections.

We begin by showing how our formalism reproduces the one-loop shift.
As already mentioned, the effective action \effectotal\ can be recast,
up to first order in perturbation theory, into the form:
$$
\eqalign{
\gm_{\rm local}\,(A,b,c,G,H;k)\,=\,
      & -\,{i(k+\cv)\over 4\pi}\,\,S_{CS}\,(A)
         - \,\idx\,\, b^a\partial A^a \,+\,\dd X_{\rm gf}\,
             +\,\dd X \crr
& + \,{\rm two\!-\!loop~contributions} ~ , \cr}
\eqn\localone
$$
with
$$
X_{\rm gf}\,=\,-\,\idx H^a c^a ~,\quad
X\,=\,{2\over 3}\,\,\yy\,\idx\, G^a_{\m} A^{a\m} ~,
$$
$\dd$ the Slavnov-Taylor operator \SToper\ and $G^a_{\m}$ as in \combination.

We already see from eq. \localone\ that the effect of first order
radiative corrections is twofold. First, there is the shift of the
classical parameter $k$ in front of the classical CS action $S_{CS}.$
This is the one-loop shift first computed by Witten [\Witten ] and by
several authors afterwards [\AlvarezG-\Semenoff]. Notice that
$S_{CS}$ is the only gauge-invariant part of the functional \localone,
which in turn generates the local cohomology of the opertor
$\dd$ upon multiplication by $c\!$-numbers. Secondly, there is the
term $\dd X.$ This term is gauge-dependent, as illustrates the fact
that it vanishes in the background field gauge [\Carmelo]
and depends on the light-cone vectors in the light-cone
gauge[\Leibbrandt ].
 The contribution $\dd X$ is both
BRS-invariant, for the operator $\dd$ annihilates it, and cohomologically
trivial with respect to $\dd.$ Since the Slavnov-Taylor operator $\dd$ can be
considered the quantum generalization of the classical BRS operator $s$
in \brstrans, one would then not marvel if $\dd X$ did not contribute
to gauge-invariant quantities. The cohomological triviality of $\dd X$
implies that there exists a wave function renormalization of the fields
$$
A^a_{\m}\,=\,Z_A\, A'^{a}_{\m} ~~ , ~~
b^a\,=\,Z_b\, b'^{a} ~~,~~
G^a_{\m}\,=\,Z_G\, G'^{a}_{\m}~~,~~
c^a\,=\,Z_c\, c'^a~~,~~
H^a\,=\,Z_H\, H'^{a}~ ,
$$
with
$$
Z_A\,=\,Z_G^{-1}\,=\,Z_b^{-1}\quad ,\quad  Z_c\,=\,Z_H^{-1} ~  ,
$$
such that if $\gm^{(0)} (\Phi,k)$ denotes
the tree-level action, the following equation holds at one loop:
$$
\dd X\,=\,\gm^{(0)} \left( Z^{-1}_{\Phi}\Phi\,,\,k\right)\,
        -\, \gm^{(0)}(\Phi,k) ~ .
$$
This equation holds for the more general $X$ and $W_{\rm local}$ of
eqs. \generalX\ and \solution, and for the values in \wone\ it yields:
$$
Z_A\,=\,Z_G^{-1}\,=\,Z_b^{-1}\,=\,1- {2\over 3}\,\yy ~~,~~
Z_c\,=\,Z_H^{-1}\,=\,1 ~ .
\eqn\waveone
$$
Indeed, this wave function renormalization transforms the action
\localone\ into
$$
\eqalign{
\gm_{\rm local}(A', b',c',G',H';k)\,=\,
       & -\,{i(k+\cv)\over 4\pi}\,\, S_{CS}(A')\crr
& + \,\idx\,\,\left[\,- b'^a\partial A'^a
        + G'^{a\m}\,D^{ab}_\m(A")\,c'^b
            - {1\over 2}\,f^{abc} H'^a c'^b c'^c\, \right]\crr
&+\,{\rm two\!-\!loop ~ contributions} ~ . \cr}
$$
We thus see that the shift in $k$ is all that remains after the
rescaling of the fields \waveone. The fact that $Z_c$ does not receive
radiative contributions is a special feature of Landau
gauges\REF\Sorella{O. Blasi, O. Piguet and S.P. Sorella, Nucl. Phys.
{\bf B356} (1991) 154.}[\Sorella].

The properties of $\dd X$ just described imply that it does not
contribute to the vacuum expectation values of Wilson loops. To prove
that this is the case lies outside the scope of this paper; it is a
technical problem that will be reported elsewhere
\REF\Carmelou{C. P. Martin, in preparation.}
[\Carmelou]. Suffice
it to say that the gauge field contributions in $\dd X$ cancel against
the large $m$ contributions coming from the integrals
$$
\eqalign{
&\qquad\qquad\quad \int_{0}^{1} dt_1\,\int_{0}^{t_1} dt_2\,\int_{0}^{t_2} dt_3~
{\dot x}^{\m}(t_1)\,{\dot x}^{\r}(t_2)\,{\dot x}^{\n}(t_3)
{}~<\! A_{\m}(t_1)\,A_{\r}(t_2)\,A_{\n}(t_3)\! > ~,\cr
&\int_{0}^{1} dt_1\,\int_{0}^{t_1} dt_2\,\int_{0}^{t_2} dt_3
\,\int_{0}^{t_3} dt_4~
{\dot x}^{\m}(t_1)\,{\dot x}^{\r}(t_2)\,{\dot x}^{\n}(t_3)
\,{\dot x}^{\sigma}(t_4)
{}~<\! A_{\m}(t_1)\,A_{\n}(t_3)\!><\!A_{\r}(t_2)A_{\sigma}(t_4)\! > ~,\cr
}
$$
wherever two gauge fields
and three gauge fields, respectively, lie arbitrarily close
to each other on
the Wilson line. The Green functions inside these line integrals
are computed for finite $m$.
We would like to stress that this cancelation is
not accidental but a consequence of BRS invariance, namely, that the
one-loop gauge-dependent contributions to $\gm_{\rm local}$
are of the cohomologically trivial form $\dd X.$
The cancelation mechanism goes beyond CS theory
\REF\Polyakov{A.M. Polyakov, Nucl. Phys.
{\bf B164} (1979) 171.\subpar J.L. Gervais and A. Neveu, Nucl. Phys.
{\bf B163} (1980) 189. \subpar  V.S. Dotsenko and S.N. Vergeles,
Nucl. Phys. {\bf B169} (1980) 527.}
and is implied by BRS invariance [\Polyakov], this
symmetry being explicitly preserved in our formalism.

Let us now consider second order radiative corrections. First, we
substract from eq. \effectotal\ two-loop corrections arising
from the one-loop contribution in $\dd X$ that enters the
two-loop diagrams through one-loop subdiagrams. This is achieved
by means of the wave function renormalization \waveone. In terms of
the renormalized fields, the local part of the effective action reads:
$$
\eqalign{
\gm_{\rm local} (A', b',c',G',&H';k) \,=\, -\,{i(k+\cv)\over 4\pi}\,\,
                                                  S_{CS}(A')\crr
&+\,\idx\,\,\left[\,- b'^a\partial A'^a + G'^{a\m}\,D^{ab}_\m(A')\,c'^b
              - {1\over 2}\,f^{abc} H'^a c'^b c'^c\, \right]\crr
&-\,c_2\, \idx\,\,\left[\, {ik\over 4\pi}\,\,\ee^{\m\r\n}
   \left( A'^a_\m \partial_\r A'^a_\n \,
      +\, {1\over 2}\,f^{abc} A'^a_\m A'^b_\r A'^c_\n\right)\,
         +\,G'^{a\m}\partial_{\m}c'^a\ \,\right] \crr
&+\,{\rm three\!-\!loop~contributions} ~ , \cr}
\eqn\oneren
$$
with
$$
c_2\,=\,{{169+L}\over 72} \,{\left(\yy\right)}^{\!2}
$$
and $L$ as in \constant. The functional \oneren\ can be recast as
$$
\eqalign{
\gm_{\rm local}\,(A',\,b',\,c',\,G',\,H';k)\,=\,
  & -\,{i(k+\cv)\over 4\pi}\,\,S_{CS}\,(A') \,
      - \,\idx\,\, b'^a\partial A'^a \,+\,\dd' X'_{\rm gf} \,
         +\,\dd' X' \crr
& + \,{\rm three\!-\!loop~contributions} ~ , \cr}
\eqn\localtwo
$$
where
$$
X'_{\rm gf}\,=\,-\,\idx H'^a c'^a ~, \quad
X'\,=\,c_2\,\yy\,\idx\, G'^a_{\m} A'^{a\m} ~ .
$$
and $\dd'$ is the Slavnov-Taylor operator in terms of
$A'^a_\m,\,\,c'^a,\,\,G'^a_\m$ and $H'^a,$ obtained from $\dd$ in eq.
\SToper\ by replacing every field with its primed counterpart.
We see in eq. \localtwo\ that, once the one-loop wave function
renormalization \waveone\ has been performed, all second order radiative
corrections fit into the cohomologically trivial functional $\dd' X'$
in the sense of the cohomology of $\dd'.$
Notice that the coefficient in front of $S_{CS}(A')$
does not receive any two-loop contribution.
Upon multiplication by $c\!$-numbers, the functional
$S_{CS}(A')$ generates the local
cohomology of $\dd'$ over the space of integrated local
functionals of mass dimension three and
ghost number zero.

The fact that the wave function renormalization \waveone\ leads to the
effective action \localtwo\ can be partially understood by using BRS
invariance. To see this we first notice that the effect of the
renormalization \waveone\ on the two-loop BRS equation \twobrs\ is to
cancel the contribution $Y_3$ of eq. \ythree\ so that eq. \twobrs\ leads
to the following equation for $\gmbt_{\rm local}(A',c',G',H';k):$
$$
\dd'\,\gmbt_{\rm local}(A',c',G',H';k)\,=\, 0 ~ .
$$
The most general solution of this equation is
$$
-\, {i\gamma\over 4\pi}\,S_{CS}(A')\,+\,\dd'\,Y'\,\, ,
$$
where $Y'$ is given by eq. \generalX\ with every field replaced with its
primed counterpart. Hence, if $\gamma$ is equal to zero, the functional
$\gmbt_{\rm local}(A',c',G',H';k)$ can not have but the form $\dd'Y'.$
This is precisely what happens in our case, and thus, eq. \localtwo\ is
in part a consequence of the BRS invariance of the renormalized action.
We say in part because the piece of information $\gamma=0$ is not provided
by BRS invariance. On the contrary, it results from our computations in
previous sections. Did $\gamma$ not vanish, there would be a two-loop
correction $\gamma$ to Witten's one-loop shift $k \to k+\cv.$

Again, the cohomological triviality of $\dd'X'$ implies that the latter
can be absorbed by the following two-loop renormalization of the fields:
$$
A'^{a\m}=Z_{A'}\, A^{a\m}_{\WW} ~ , ~
b'^a=Z_{b'}\, b^{a}_{\WW} ~ , ~
G'^{a\m}=Z_{G'}\, G^{a\m}_{\WW} ~ , ~
c'^a=Z_{c'}\, c^a_{\WW} ~ , ~
H'^a=Z_{H'}\, H^{a}_{\WW} ~ ,
$$
with
$$
Z_{A'}\,=\,Z_{G'}^{-1}\,=\,Z_{b'}^{-1}\,=\,
      1\,-\,{{169+L}\over 72} \,{\left(\yy\right)}^{\!2}\
{}~ , \quad  Z_{c'}\,=\,Z_{H'}^{-1}\,=\,1 ~ .
$$
We also have that if $\gm^{(0)}(\Phi',k)$ is the tree-level
action in terms of the fields $A'^a_\m,\,\,b'^a,\,\,c'^a,\,\,G'^a_\m$
and $H'^a,$ the identity
$$
\dd' X'\,=\,\gm^{(0)} \left( Z^{-1}_{\Phi'}\Phi'\,,\, k\right)\,
         -\, \gm^{(0)}(\Phi' , k)\,\
$$
holds at the order we are working. The local part of the effective
action in terms of the fields
$A^{a\m}_{\WW},\,b^a_{\WW},\,c^a_{\WW},\,G^{a\m}_{\WW}$ and $H^a_{\WW}$
takes the form:
$$
\eqalign{
\gm_{\rm local}(A_{\WW},b_{\WW},c_{\WW},G_{\WW},H_{\WW};k) \,=\,
                     & -\,{i(k+\cv)\over 4\pi}\,\,S_{CS}(A_{\WW})\crr
&+\,\idx\,\left[\,- b_{\WW}^a\partial A_{\WW}^a \,
     +\,G_{\WW}^{a\m}\,D^{ab}_\m(A_{\WW})\,c_{\WW}^b\,
       -\,{1\over 2}\,f^{abc}\,H_{\WW}^a c_{\WW}^b c_{\WW}^c \,\right]\crr
&+\, {\rm three\!-\!loop~ contributions} ~ . \cr}
\eqn\renormalized
$$
The parameter in front of the gauge-invariant part $S_{CS}$ is still
the one-loop parameter $k+\cv.$ This shows that there are no
second order radiative corrections to the one-loop shift of the classical
parameter $k.$ In other words, local two-loop corrections in the
bare effective action \effectotal\ correspond to a wave function
renormalization and are, therefore, unobservable [\Collins]. The
existence of such renormalization follows from the cohomologically
trivial form $\dd' X'$  in which the two-loop contributions can
be written and can be seen as a
consequence of BRS invariance. The rationale why $\dd' X'$ should not
contribute to the vacuum expectation values of Wilson loops is very much
the same as for $\dd X,$ roughly scketched above. The outcome of this
discussion is that still after taking into account second order radiative
corrections, the monodromy parameter is $k+\cv.$ This result is in
agreement with Wilson loop computations for the unknot
[\Alvarez], which demand that two-loop corrections to
the vacuum polarization tensor do not contribute to the Wilson loop, if
Witten's exact result is to be recovered in perturbation theory.

We finish this section by collecting the wave function
renormalization that transforms the local effective
action \effectotal\ into the form \renormalized:
$$
\eqalign{
&Z_b^{-1}\,=\,Z_G^{-1}\,=\,Z_A \,=\, 1 - {2\over 3}\, \yy  -
   \left( {169 \over 72} + {22 \over 3}\, \ln 2
       - {63 \over 8}\,\ln 3 \right) {\left(\yy\right) }^2 ~ .\crr
&Z_c\,=\,Z_H^{-1}\,=\,1\,\, .\cr}
$$
Both actions \effectotal\ and \renormalized\ yield the same
values for the observables of the theory. However, the effective
action \renormalized\ clearly displays what the radiative
corrections contributing to the observables are.

{\bf\chapter{Conclusions}}

Working in the covariant Landau gauge, we have studied
CS theory as the large mass limit of TMYM theory, the latter
being defined as the standard Yang-Mills theory in
three dimensions with a CS term responsible for a topological mass
$m.$ The idea behind this approach is to use the Yang-Mills term as
a higher-covariant derivative regularization so that CS quantities
are computed by sending $m$ to infinity in their TMYM counterparts.
However, this is not good enough, since TMYM theory is not finite
by power counting and, hence, the
large $m$ limit can not be taken directly. One first has to
introduce an additional regulator to cure TMYM divergences.
Here we have used dimensional regularization and have
shown that it is BRS-invariant. Our approach to CS theory
can then be viewed as a hybrid BRS-invariant regularization consisting
of a higher covaraint derivative Yang-Mills term plus dimensional
regularization. In particular, bare CS Green functions are defined as
the limit $m\to \infty$ of the limit $D\to 3$ of the corresponding
dimensionally regularized TMYM Green functions.

We have seen that the limit $D \to 3$ of any dimensionally
regularized TMYM Green function is well-defined at any perturbative
order. This shows that TMYM theory is perturbatively finite and guarantees
that taking the limit $m\to\infty$ is legitimate. To compute the latter
limit we have given two large-mass vanishing theorems. One is thus
left with all necessary tools to calculate. In this
paper we have computed the local part of the CS effective action up to
second order in perturbation theory. We have done this by solving, up
to two loops, the BRS identities for the effective action over the
space of local functionals. The coefficients entering in
the solution can be determined by exploiting that the effective action
generates 1PI Green functions. This demands computing three 1PI Green
functions. Here we have chosen the vacuum polarization tensor, the
ghost self-energy and the ghost-ghost-external field vertex. Their
one- and two-loop expressions are
given in \oneloop , \twoloop , \ghostone\ -\extone\ and \exttwo ,
which in turn yield for the bare effective action eq. \effectotal .

To construct the quantum theory, regularization is not enough but
has to be supplemented with renormalization conditions. Notice that
the theory being finite does not necessarily imply that an UV finite
renormalization is not needed.
As a matter of fact, there are many finite renormalizations
leading to different expressions for the renormalized effective action
and the observables (see
[\GMR] for examples). Of course, for a finite theory like ours,
one can always choose a renormalization scheme in which the renormalized
coupling constant and fields are equal to the bare ones, but this is already
a choice. The problem is that there are no arguments within the framework of
local and renormalized perturbative quantum field theory that lead naturally
to a renormalization scheme. The reasons for a given parametrization have
to be found outside this framework. For finite theories one may assume that
the classical parameters constitute the right parametrization, provided
a regularization preserving the symmetries of the observables or fundamental
symmetries is used\REF\Jaffe{A. Jaffe and A. Lesniewski, {\it Supersymmetric
quantum fields and infinite dimensional analysis}, in {\it Non-perturbative
quantum field theory}, G. 't Hooft, A. Jaffe, G. Mack P.K. Mitter and R.
Stora eds., Plenum Press (1988)}[\Jaffe]. The idea is that the fundamental
symmetries, when respected at the regularized level, are strong enough
as to fix all ambiguities with observables effects introduced by
regularization. For this parametrization to make sense, the emerging quantum
theory has to be independent of the particular invariant regularization
one uses; or in other words, all regularizations preserving the fundamental
symmetries have to give the same values for the observables as functions
of the bare parameters. In the case at hand, the only symmetry of the
observables is gauge invariance, which in the gauge-fixed theory becomes
BRS invariance. Furthermore, it is known that all BRS-invariant
regularizations of CS theory give the same result for the expectation values
of Wilson loops in terms of the bare or classical $k,$ at least at first order
in perturbation theory [\AlvarezG-\Semenoff]. In addition,
also at this order, the values in terms of the bare $k$ for the
monodromy parameter of the observables and for the central extension of the
2-dimensional current algebra are the same as for the hamiltonian formalism
[\AlvarezGtwo]. Based upon these two facts we have parametrized the theory in
terms of the bare or classical $k.$

With this choice of parametrization we have finally analyzed the
relevance of the radiative corrections we have computed.
In gauge theories, only gauge-invariant (rather than merely BRS-invariant)
corrections contribute to the observables.
To extract gauge-invariant radiative corrections from
the local effective action, we have studied the cohomology of
the Slavnov-Taylor operator, since contributions cohomologically trivial
do not contribute to the observables. As a result, we have obtained
for the effective action that:

{\leftskip=1 true cm  \rightskip=1 true cm
\noindent {\it i}) Part of one-loop radiative corrections are
gauge-invariant and produce the shift $k \to k +\cv$ of the bare CS
parameter $k,$ in agreement with previous results in the literature
[\Witten,\AlvarezG-\Semenoff]. The remaining one-loop
contributions are gauge-dependent. Indeed, they are cohomologically trivial
with respect to the Slavnov-Taylor operator so that they can be set to zero
by means of a finite multiplicative wave function renormalization of
only the fields. Hence, they do not contribute to the observables of the
theory [\Carmelou], in agreement with the general statement that
UV finite renormalizations of the fields are unobservable [\Collins].

\noindent {\it ii}) All two-loop radiative corrections are
cohomologically trivial in the sense of the Slavnov-Taylor operator,
hence gauge-dependent. They can thus be set to zero by renormalizing
the fields. Two-loop corrections to the effective action do not contribute
therefore to the observables of the theory. This means that the
shift $k \to k +\cv$ does not receive second order contributions.
\par}

We would like to mention that the analysis presented
here clearly displays what the options are. If one parametrizes the
quantum theory in terms of the classical $k,$ the one-loop shift
$k \to k +\cv$ does not receive two-loop contributions.
If, on the other hand, one insists in a renormalized CS parameter $k_{\WW},$
it does not make any sense to speak of a shift. Obviously, one can tunnel
between these two situations by a finite renormalization $k_{\WW}= k +\cv.$
We think, however, that the fact that all BRS-invariant regularizations
tried in the literature give the same shift in $k$ when the quantum
theory is parametrized in terms of the classical CS parameter is no accident,
and that there must be a fundamental reason for that. Let us stress that
the latter does not mean that the one-loop Landau-gauge bare effective action
should be the same for all BRS-invariant regularizations, since the
effective action is not an observable itself. It rather means that
the corrections in the effective action that contribute to the
observables of the theory should be the same for all BRS-invariant
regularizations. This is what one
actually observes in the existing literature: gauge-invariant,
or equivalently cohomologically non-trivial one-loop corrections to the
effective action are the same for any BRS-preserving regularization
and provide the famous shift. The remaining one-loop corrections
to the bare effective action depend on the BRS-invariant regularization
one uses. However, they fit into a cohomologically trivial functional,
\ie\ they correspond to wave function renormalizations of only the fields
and are therefore unobservable. To summarize, bare Green functions provided
by different BRS-invariant regulators differ as much as possible in so far
as the shift of the classical parameter remains unchanged. According to
the general principles of local and renormalized perturbation theory
[\Hepp], these differences should even reach the shift, unless there is
some reason why they should not.

Let us close by anticipating what three-loop radiative
corrections should be, if the pattern we have unveiled occurs beyond
two loops. Once one- and two-loop cohomologically trivial contributions
have been absorbed by appropriate wave function renormalizations, all
three-loop corrections should thoroughly fit into a cohomologically trivial
functional with respect to the renormalized Slavnov-Taylor operator.
The value of these corrections will in general depend on the particular
BRS-invariant regulator used. There should be however no contributions to
the one-loop shift.

\medskip

\ack

The authors are grateful to Professors G. Leibbrandt and J.C. Taylor
for valuable conversations and encouragement.
GG was supported by the Italian Consiglio Nazionale delle
Ricerche, CPM by the National Science and
Engineering Research Council of Canada under grant No.
A-8063, and FRR by The Commission of the European Communities
through contract No. SC1000488 with The Niels Bohr Institute.
Their work also benefited from partial support from CICYT, Spain.

\vfil\eject

{\bf\Appendix{}}

We give here a list of integrals that have to be computed
to obtain the explicit expressions for the gauge field polarization
tensor and the ghost self-energy of eqs. \twoloop\
and \ghosttwo. As explained in Sect. 5, the use of some algebraic
identities and of the $m\!$- and $o\!$-theorems reduces the problem to
that of calculating a finite number of one- and two-loops integrals.
One-loop integrals are of the form
$$
\int {{d^D\!q}\over{(2\pi)^D}}\,\,
   {1 \over{{\left( q^2 \right)}^{\a}\,
            {\left( q^2+1 \right)}^{\b}}} \,=\,
{{\gm\!\left(\dhalf-\a \right)\,\gm\!\left(\a+\b-\dhalf\right)}
   \over { \pidhalf \, \gmbe \,\gmdhalf }} \,\, .
$$
As for the two-loop integrals they can be classified into
four types, namely:
$$\eqalign{
I_{\a\b\ga}^{(0)} \,&=\, \meas
     {1\over {{\left( k^2 \right)}^{\a}\,
         {\left( q^2 \right)}^{\b}\,
             {\left[ (k+q)^2 \right]}^{\ga}}} \,\, , \cr
I_{\a\b\ga}^{(1)} \,&=\, \meas
     {1\over {{\left( k^2+1 \right)}^{\a}\,
         {\left( q^2 \right)}^{\b}\,
             {\left[ (k+q)^2 \right]}^{\ga}}} \,\, , \cr
I_{\a\b\ga}^{(2)} \,&=\, \meas
     {1\over {{\left( k^2+1 \right)}^{\a}\,
         {\left( q^2+1 \right)}^{\b}\,
             {\left[ (k+q)^2 \right]}^{\ga}}} \,\, , \cr
I_{\a\b\ga}^{(3)} \,&=\, \meas
     {1\over {{\left( k^2+1 \right)}^{\a}\,
         {\left( q^2+1 \right)}^{\b}\,
             {\left[ (k+q)^2+1 \right]}^{\ga}}} \,\, . \cr
}$$

In dimensional regularization integrals of the first type vanish
$$
I_{\a\b\ga}^{(0)}=0 \,\, .
$$

To evaluate $I_{\a\b\ga}^{(1)}$ and $I_{\a\b\ga}^{(2)}$ we
use the method of Feynman parameters. It is straightforward to
show that
$$
I_{\a\b\ga}^{(1)}\,=\,
   {{\gm\!\left(\dhalf-\b\right)\,\gm\!\left(\dhalf-\ga\right)\,
        \gm\!\left(\b+\ga-\dhalf\right)\,
             \gm (\a+\b+\ga-D)} \over
    {\pid \,\, \gmal \,\, \gmbe \,\, \gmga \,\, \gmdhalf}} \,\, .
$$
The computation of $I_{\a\b\ga}^{(2)}$ is however more involved.
We next present some of its details. Integrating over $q$ we
obtain:
$$
\eqalign{
I_{\a\b\ga}^{(2)}\,=\, {{\gm\!\left(\ga+\b-\dhalf\right)}
             \over {\pidhalf \,\, \gmga \,\, \gmbe}}
     & \int {{d^D\!k}\over{(2\pi)^D}}\,\,
             {\left( k^2+1 \right)}^{\dhalf-\a-\b-\ga} \cr
{\scriptstyle \times} & \int_0^1 dx \,\, x^{\dhalf-\ga-1}
     \,\, (1-x)^{\ga-1}\,\,
        \left( 1 - {k^2 \over {k^2+1}}\, x\right)^{\dhalf-\ga-\b}
\,\, . \cr}
$$
Integration over $x$ gives \REF\grads{I. S. Gradshteyn and I. M.
Ryzhik, {\it Table of integrals, series and products}, XX edition,
Academic Press (1980).}[\grads]:
$$
B\!\left({{\scriptstyle D}\over{\scriptstyle 2}}-\ga
       \, , \, \ga\right) \,
  F\!\left( \ga+\b-{{\scriptstyle D}\over{\scriptstyle 2}}\, , \,
   {{\scriptstyle D}\over{\scriptstyle 2}} - \ga \, ; \,
     {{\scriptstyle D}\over{\scriptstyle 2}} \, , \,
        {{\scriptstyle k^2} \over {\scriptstyle k^2+1}} \right) ~,
$$
where $B$ is Euler's beta function and $F$ is Gauss' hypergeometric
function. To integrate over $k$ we use polar coordinates and
perform the change of variables $t={\textstyle {k^2\over {k^2+1}}}.$
In this way we get
$$
\eqalign{
I_{\a\b\ga}^{(2)}\,=\,\,&
    {{\gm\!\left(\ga+\b-\dhalf\right) \,
        \gm\!\left( \dhalf-\ga \right) }
            \over {\pid \,\, \gmbe\,\,
                {\left[ \gm\!\left( \dhalf \right)\right]}^2}}
                                                   \cropen{12pt}
&\qquad {\scriptstyle \times}
    \int_0^1 dt ~ t^{\dhalf -1} ~ (1-t)^{\a+\b+\ga-D-1} ~
      F\!\left( \ga+\b-{{\scriptstyle D}\over{\scriptstyle 2}}\, , \,
        {{\scriptstyle D}\over{\scriptstyle 2}} - \ga \, ; \,
          {{\scriptstyle D}\over{\scriptstyle 2}} \, , \, t \right) ~.
\cr}
$$
The integral on the RHS can be performed [\grads] and gives a
certain product of Euler's gamma functions so that
$$
I_{\a\b\ga}^{(2)}\,=\, {{\gm\!\left( \ga + \b - \dhalf \right) \,
     \gm\!\left( \dhalf-\ga \right) \,
     \gm (\a+\b+\ga-D) \,\, \gm\!\left( \a+\ga - \dhalf\right)}
     \over {\pid \,\, \gmal \,\, \gmbe \,\,
                \gm (\a+\b+2\ga-D) \,\, \gmdhalf }} \,\, .
$$
It should be stressed that all these manipulations are
in the sense of dimensional regularization. More precisely, they
hold for $D,\,\,\a,\,\,\b$ and $\ga$ in suitable domains. To
obtain the value of $I_{\a\b\ga}^{(2)}$ outside them one uses
analytic continuation.
This same observation applies to all the integrals in this appendix.

For $I_{\a\b\ga}^{(3)}$ we have not found
a compact expression in terms of the parameters
$D,\a , \b$ and $\ga.$ However, the result
does look very simple for specific values
of them. Here we list all integrals of this type at $D=3$
encountered when performing the calculations
of Sects. 5 and 6:
$$
I_{111}^{(3)}\,=\,
 {1 \over {32\pi^2}} \left\{ -\,\,{1\over{\eps}} +
     1 - \ga + \ln\left({4\pi \over 9}\right)
        + O(\eps) \right\} \,\, ,
$$
$$
I_{121}^{(3)}\,=\, {1\over{96\pi^2}} \quad , \quad
I_{131}^{(3)}\,=\, {1\over{288\pi^2}} \quad , \quad
I_{221}^{(3)}\,=\, {1\over{576\pi^2}} \quad , \quad
I_{231}^{(3)}\,=\, {5\over{6912\pi^2}} \,\, ,
$$
where $\eps=D-3$ is the regulator.

Note that $I_{111}^{(1)}\, , \, I_{111}^{(2)}$ and $I_{111}^{(3)}$
are overall UV divergent at $D=3,$ exhibiting a pole at $\eps=0.$

\vfil\eject

\refout

\vfil\eject

{\centerline {\fourteenbf Figures' captions}}

\bigskip

Fig. 1: Feynaman rules for the action \maction.

\medskip

Fig. 2: One-loop diagrams contributing to the vacuum polarization tensor.

\medskip

Figs. 3, 4, 5 and 6: Two-loop diagrams contributing to the vacuum
polarization tensor.

\medskip

Fig. 7: One-loop diagram contributing to the ghost self-energy.

\medskip

Fig. 8: Two-loop diagrams contributing to the ghost self-energy.

\medskip

Fig. 9: One-loop diagrams contributing to the ghost-ghost-external field
vertex.

\medskip

Fig. 10: Two-loop diagrams contributing to the ghost-ghost-external field
vertex.

\bye